\documentstyle[prd,aps]{revtex}

\begin{document}

\title{Facets of Tunneling: Particle production in external fields} 
\author{K.~Srinivasan\thanks{Electrorinc address:~srini@iucaa.ernet.in},  
T.~Padmanabhan\thanks{Electronic address:~paddy@iucaa.ernet.in}}
\address{IUCAA, Post Bag 4, Ganeshkhind, Pune 411 007, INDIA.}

\maketitle

\begin{abstract}
This paper presents a critical review of particle production in an 
uniform electric field and Schwarzchild-like spacetimes. 
Both problems can be reduced to solving an effective one-dimensional 
Schrodinger equation with a potential barrier.  
In the electric field case, the potential is that of an inverted 
oscillator $-x^2$ while in the case of Schwarchild-like spacetimes, 
the potential is of the form $-1/x^2$ near the horizon.  
The transmission and reflection coefficients can easily be obtained for 
both potentials. 
To describe particle production, these coefficients have to be 
suitably interpreted. 
In the case of the electric field, the standard Bogoliubov coefficients 
can be identified and the standard gauge invariant result is recovered. 
However, for Schwarzchild-like spacetimes, such a tunnelling 
interpretation appears to be invalid.  
The Bogoliubov coefficients cannot be determined by using an 
identification process similar to that invoked in the case of the 
electric field. 
The reason for such a discrepancy appears to be that, in the tunnelling 
method, the effective potential near the horizon is singular and 
{\it symmetric}.  
We also provide a new and simple semi-classical method of obtaining 
Hawking's result in the $(t,r)$ co-ordinate system of the usual 
standard Schwarzchild metric. 
We give a prescription whereby the singularity at the horizon can 
be regularised with Hawking's result being recovered.  
This regularisation prescription contains a fundamental asymmetry 
that renders both sides of the horizon dissimilar.  
Finally, we attempt to interpret particle production by the electric 
field as a tunnelling process between the two sectors of the 
Rindler metric.    
\end{abstract}



\section{Introduction and Summary}
In this paper, we present a critical review of particle production 
in Schwarzchild-like spacetimes and in an uniform electric field in 
Minkowski spacetime.   
Our emphasis here is on the tunnelling picture used to interpret 
results obtained by other (presumably more reliable!) methods. 

Consider the problem of a scalar field propagating in flat spacetime 
in an uniform electric field background. 
The total particle production due to the presence of the electric 
field upto the one-loop approximation is correctly calculated by 
the gauge invariant method proposed by Schwinger~\cite{schwinger}. 
The same problem can be reduced, in a time dependent gauge, to an 
equivalent Schrodinger equation with an inverted harmonic oscillator 
potential.
Such an equation can be solved exactly using the standard flat 
spacetime quantum field theoretic methods. Since the problem is 
explicitly time dependent, the vacuum state at $t \to -\infty$ and 
at $t \to \infty$ are not the same. 
The Bogoliubov coefficients between the ``in'' and ``out'' vacua 
are easily calculated and the total particle production turns out 
to be the same as calulated by the Schwinger method.  
However, if a space dependent gauge is used to describe the same field, 
the vacuum state of the field remains the same for all time and hence 
no particle production can take place.  
To recover the standard result, the tunnelling interpretation is 
introduced. This interpretation is useful since it provides a 
dynamical picture of particle production (see, for example \cite{brout95}) 
and is the only way in which the standard gauge invariant result 
can be recovered in a time independent gauge.
In this paper, we attempt a tunnelling description for both 
time dependent and time independent gauges. 
The method of complex paths, enunciated by Landau in~\cite{landau3}, 
is used to calculate the transmission and reflection coefficients for 
the equivalent quantum mechanical problem.  
Then,  an interpretation of these coefficients, in order to obtain 
the standard gauge invariant result is provided.   

Let us next consider Schwarzchild-like spacetimes like the usual 
black hole spacetime, the Rindler spactime and the de-Sitter spacetime. 
In the standard black-hole spacetime, particle production was 
obtained by Hartle and Hawking~\cite{hawking76} using 
semi-classical analysis.
In this method, the semi-classical propagator for a scalar 
field propagating in the Schwarzchild spacetime is analytically 
continued in the time variable $t$ to complex values.  
This continuation gives the result that the probability of 
emission of particles from the past horizon is not the same 
as the probability of absorption into the future horizon. 
The ratio between these probabilities is of the form
\begin{equation}
P[{\rm emission}] = P[{\rm absorption}] e^{-\beta E}
\end{equation}
where $E$ is the energy of the particles and $\beta = 1/8\pi M$ 
is the standard Hawking temperature. 
The above relation is interpreted to be equivalent to a thermal 
distribution of particles in analogy with that observed in an 
atom interacting with black body radiation. 
In the latter case,  the probability of emission of radiation by 
the atom is related to the probability of absorption by the atom 
by a similar relation as given above. 
Now, in Hawking's derivation, the Kruskal extension is of vital 
importance in obtaining the thermal spectrum. 
In this  paper, we propose an alternate derivation of Hawking 
radiation without using the Kruskal extension.  
We consider the semi-classical propagator in the standard Schwarzchild
co-ordinates near the horizon and then use a prescription to bypass 
the singularity there.  
This prescription gives the same result as that obtained by Hawking 
and can be used in all spacetimes with a Schwarzchild-like metric.
We then attempt to create a tunnelling picture on the lines of 
that done for the electric field.  
The equivalent Schrodinger potential in this case is of the 
form $-1/x^2$ near the horizon.  
The transmission and reflection coefficients can easily be 
obtained using Landau's method of complex paths. To obtain particle 
production, these coefficients have to be suitably interpreted. 
In the case of the electric field, the standard Bogoliubov 
coefficients can be identified by recasting the normalisation 
rule between the transmission and reflection coefficients, 
namely $T+R =1$, in such a way that the standard gauge invariant 
effective lagrangian result is recovered. 
However, in the case of the Schwarzchild-like spacetimes, 
such a tunnelling interpretation appears to be invalid.  
The Bogoliubov coefficients cannot be identified using the 
transmission and reflection coefficients in the same way as was 
done in the case of the electric field since the spectrum of 
created particles is then not thermal as Hawking's semiclassical method 
suggests.  
But reinterpreting these coefficients also does not give the 
standard result. 
The reason for such a discrepancy appears to be that in the 
tunnelling method, the effective potential near the horizon 
is symmetric and singular and hence both sides of the horizon 
appear to be the same.

Finally, we attempt to link particle production by an uniform 
electric field with processes occuring in the Rindler frame.  
We propose that tunnelling between the two Rindler sectors 
gives rise to the production of real particles.
We do this entirely in a heuristic manner and show that this 
tunnelling process between the two Rindler sectors gives 
rise to the exponential factors in the expression for the 
effective lagrangian (see Eqn.~(\ref{eqn:efflag}). 
But the coefficients multiplying the exponential factors, however, 
must be determined by solving the problem exactly.

The layout of the paper is as follows. 
In section~(\ref{sec:facets}), we will discuss first the 
calculation of the reflection and transmission coefficients 
using Landau's powerful method of complex paths for the 
potentials $-x^2$ and $-1/x^2$.  
Then, in section~(\ref{sec:electric}) we apply the results of 
the previous section to particle production in an electric field. 
In section~(\ref{sec:schwarzchild}), the semi-classical derivation 
of Hawking radiation without taking recourse to the Kruskal 
extension will be extensively discussed. 
Next, the reduction of the problem to an effective Schrodinger 
equation and tunnelling at the horizon will be considered.     
Finally, in section~(\ref{sec:cpaths}), we attempt to link 
particle production in the electric field to processes occuring 
in the Rindler frame.

\section{Facets of Tunneling} \label{sec:facets}
In this section we briefly review the basic concepts of 
semi-classical quantum mechanics in one dimension and 
formally describe the tunnelling process.  
We then apply the formalism to two potentials, namely, 
$V_1(x) = -x^2$ and $V_2(x) = -1/x^2$, and calculate the 
transmission and reflection coefficients for both.  

\subsection{Semi-classical limit of Quantum Mechanics} 
\label{subsec:semiQM}
Consider a simple one dimensional quantum mechanical system with 
an arbitrary potential $V(x)$ where $x$ denotes the space 
variable (see Ref.~\cite{landau3} for details). 
To describe the transition of the system from one state to another, 
we first solve the corresponding classical equations of motion 
and determine the path of transition.  
This path is, in general, complex since many processes like 
tunneling through a potential barrier cannot occur classically.  
Therefore, the transition point $q_0$ where the system formally 
makes the transition is a complex number  determined by the 
classical conservation laws.  
Then, the action $S$ for the transition from some initial state 
$x_i$ to a final state $x_f$ given by
\begin{equation}
S(x_f, x_i) = S(x_f, q_0) + S(q_0, x_i)
\end{equation}
is calculated.  
Here, $S(q_0, x_i)$ is the action for the system to move from the 
initial state $x_i$ to the transition point $q_0$ while 
$S(x_f, q_0)$ is that to move from $q_0$ to $x_f$.  
The probability $P$ for the transition to occur is given by the formula 
\begin{equation}
P \sim \exp \left( -{2\over \hbar} {\rm Im}[S(x_f, q_0) + 
S(q_0, x_i)]\right) .
\label{eqn:semiclas1}
\end{equation}
The above formula is valid only when the exponent of the 
exponential is large.  
Further, if the potential energy has singular points, these 
must also be considered as possible values for $q_0$.
If the position of the transition point is not unique, then it 
must be chosen so that the exponent in Eqn.~(\ref{eqn:semiclas1}) 
has the smallest absolute value but still must be large enough 
so that the above formula be valid.  

If the transition point $q_0$ is real, but lies in the classically 
inaccessible region, then the above formula gives the transmission 
coefficient for penetration through a potential barrier, while if 
the transition point is complex, it solves the problem finding 
the over the barrier reflection coefficient. 
The $\sim$ sign in the above formula is used since the coefficient 
in front of the exponential is not determined. 
This can be determined by calculating the exact semi-classical wave 
functions.  
Generally, it is desirable to find the ratios of two different 
transitions so that this coefficient does not matter. 

The physics of the tunnelling and the ``over the barrier'' 
reflection processes are very different.  
In the tunnelling process, the semi-classical analysis gives a 
transmission coefficient that is an exponentially small quantity 
with the corresponding reflection coefficient being unity.  
In contrast, in the ``over the barrier'' reflection process, 
 just the reverse is obtained.  
The transmission coefficient is unity while the reflection 
coefficient is an exponentially small quantity. 
Both these processes will be encountered when the electric 
field is studied in different gauges. 

We will now review the method of calculating the transmission and 
reflection coefficients for a typical quantum mechanical problem 
using the method of complex paths for a general potential $V(x)$.  

\subsubsection{Description of the method of complex paths} 
\label{subsubsec:tunnel}

Consider the motion of a particle of mass $m$ in a region 
characterised by the presence of a potential $V(x)$ in one 
space dimension. 
The problem is to calculate the transmission and reflection 
coefficients between two asymptotic regions labelled $L$ and 
$R$ where the semi-classical approximation to the exact wave 
function is valid.  
After identifying these regions and writing down the 
semi-classical wave functions, definite boundary conditions 
are imposed.  
The usual boundary conditions considered  are that in one 
region, say $L$,  the wavefunction is a superposition of an 
incident wave and a reflected wave while in the second region 
$R$, the wave function is just a transmitted wave.  
Then, a complex path (in the plane of the now complex 
variable $x$) is identified from $R$ to $L$ such that  
(a) all along the path the semiclassical ansatz is valid 
and (b) the reflected wave is exponentially greater than 
the incident wave at least in the latter part of the path 
near the region $L$.  
The transmitted wave is then moved along the path  to obtain 
the reflected wave and thus the amplitude of reflection 
is identified in terms of the transmission amplitude.  
Having done this, the normalisation condition is used {\it i.e.} 
the sum of the modulus square of the transmission and reflection 
amplitudes should equal unity, to determine the exact values 
of the transmission and reflection coefficients.   

For a given potential, the turning points $q_0$ (or transition 
points) are given by solving the equation
\begin{equation}
p(q_0) = \sqrt{2m(E - V(q_0))} = 0 \label{eqn:tpoint1}
\end{equation}
where $p(x)$ is the classical momentum of the particle and 
$E$ is the energy of the particle. 
In general, $q_0$ is complex.  
At these points, the semi-classical ansatz is not valid since 
the momentum is zero. 
Further, the potential can possess singularities. 
At these points too, the semi-classical approximation is invalid.  
Therefore the contour between the two regions should be chosen to 
be far away from such points. 
In general the contour will enclose them.  
Therefore, the relation between the transmission and reflection 
amplitudes is determined by taking into account the turning points 
and the singularities of the potential.     

The Schrodinger equation to determine the wave function $\psi$ 
of the particle is
\begin{equation}
-{\hbar^2 \over 2m} {d^2 \psi \over dx^2}= (E - V(x))\psi . 
\label{eqn:schrodinger}
\end{equation}
Referring to~\cite{landau3}, the  semi-classical wave 
function, in the classically allowed region where $E>V(x)$, 
is given by the formula
\begin{equation}
\psi = C_1 p^{-1/2} \exp\left( {i\over \hbar} 
\int \! p(x)dx \right) + C_2 p^{-1/2} \exp\left(-{i\over \hbar} 
\int \! p(x)dx \right) 
\label{eqn:tunnel1}
\end{equation}
while in the classically inaccessible regions of space where 
$E<V(x)$, the function $p(x)$ is purely imaginary and the wave 
function is now given by the relation
\begin{equation}
\psi = C_1 |p|^{-1/2} \exp\left( -{1\over \hbar} 
\int \! |p(x)|dx \right) \;+\; C_2 |p|^{-1/2} \exp\left({1\over \hbar} 
\int \! |p(x)|dx \right) .
\label{eqn:tunnel2}
\end{equation}
The condition on the potential for semi-classicality of the 
wave function to be valid is 
\begin{equation}
\left| {d \over dx}\left({\hbar \over p(x)}\right) \right| \ll 1, 
\label{eqn:cond1}
\end{equation}
or, in another form,
\begin{equation}
{m\hbar |F| \over |p|^3} \ll 1, \qquad F = -{dV \over dx}. 
\label{eqn:cond2}
\end{equation}
It ought to be noted that the accuracy of the semi-classical 
approximation is not such as to allow the superposition of 
exponentially small terms over exponentially large ones. 
Therefore, it is inapplicable in general to retain both terms 
in Eqn.~(\ref{eqn:tunnel2}).  
We will consider a  few cases of interest in this paper and  
refer the reader to \cite{landau3} for a exhaustive discussion 
along with suitable illustrative examples.

Consider the case in which the semi-classical 
condition~(\ref{eqn:cond2}) holds in the regions 
$x\to \pm\infty$. 
As $x\to -\infty$, the wave function is assumed to be a 
superposition of incident and reflected waves and is written 
in the form, 
\begin{equation}
\psi =  p^{-1/2} \exp\left( {i\over \hbar} 
\int \! p(x)dx \right) + C_2 p^{-1/2} \exp\left(-{i\over \hbar} 
\int \! p(x)dx \right) 
\label{eqn:tunnel10}
\end{equation}
where the incident wave has unit amplitude while the 
reflected wave has amplitude given by $C_2$. As $x\to +\infty$, 
the wave function is assumed to be a  right moving travelling wave, 
\begin{equation}
\psi = {C_3 \over \sqrt{p}} \exp \left( {i \over \hbar} 
\int \! p dx  \right)  
\label{eqn:tunnel11}
\end{equation}
The method of complex paths can now be  applied on the 
function~(\ref{eqn:tunnel11}).  The contour is chosen either 
in the upper or lower complex plane such that the reflected 
wave is always exponentially greater than the incident wave 
along that part of the path near the region $x \to -\infty$.  
If this is satisfied along one of the contours then $C_2$ 
is determined in terms of $C_3$. To carry out the above 
procedure however, the exact semi-classical wave functions 
as $x\to \pm \infty$ have to be determined. 
This will be done explicitly later for the relevant cases. 

A different case arises when the semi-classical ansatz holds 
in the vicinity of the origin $x=0$ rather than at $x=\pm \infty$. 
The boundary conditions assumed in this case are the same 
as above with the condition $x \to \infty$ replaced by 
$x \to 0^+$ and $x \to -\infty$ by $x \to 0^-$. 
Here, the required contour is about the origin and is chosen 
to be small.  
But it must still be large enough so that the reflected 
wave is much larger than the incident wave along the latter 
part of the contour near the region $x < 0$.  

In the above cases the method of complex paths gives the 
exact transmission and reflection amplitudes.  
But, in certain cases it is enough to assume that the 
transmission and incident amplitudes are equal to unity 
while the reflection amplitude is exponentialy damped and 
consequently very small. 
Here, the ``over the barrier'' reflection coefficient for 
energies large enough so that the reflection coefficient is 
exponentially small, has to be determined. 
In this case, the condition $E>V(x)$ is always satisfied.  
Therefore, the transition point $q_0$ at which the particle 
reverses its direction is the complex root of the equation 
$V(q_0) = E$.  
Let $q_0$ lie in the upper complex plane for definiteness. 
Now, the amplitudes of the incident and transmitted waves 
are equal (both are set to unity within exponential accuracy).  
To calculate the reflection coefficient, the relation between 
the wave functions far to the right of the barrier and far 
to the left of the barrier must be determined.  
The transmitted wave can be written in the form
\begin{equation}
\psi_T = {1 \over \sqrt{p}} \exp\left( {i \over \hbar} 
\int_{x_1}^{x} \! p dx \right) 
\label{eqn:aboveb1}
\end{equation}
where $x_1$ is any point on the real axis.  
We follow the variation of $\psi_T$ along a path $C$ in the upper 
complex plane which encloses the turning point $q_0$.   
The latter part of this path must lie far enough to the left 
of $q_0$ so that the error in the semi-classical incident wave 
is less than the required small reflected wave. 
Passage around $q_0$ only causes a change in the sign of the 
root $\sqrt{E -V(x)}$ and after returning to the real axis, 
the function $\psi_T$ becomes the reflected wave $\psi_R$. 
Going around a complex path in the lower complex plane 
converts $\psi_T$ into the incident wave.  
Since the amplitudes of the incident and transmitted waves may 
be regarded as equal, the required reflection coefficient 
is given by 
\begin{equation}
R = \left|{\psi_R \over \psi_T} \right|^2 = 
\exp\left( - {2 \over \hbar}{\rm Im}\int_{C} \! pdx \right) 
\label{eqn:aboveb2}
\end{equation}
Now we can deform the contour in any way provided it still 
encloses the point $q_0$.  
In particular, the contour can be deformed to go from 
$x_1$ to $q_0$ and back.  
This gives
\begin{equation}
R = \exp\left( - {4 \over \hbar}{\rm Im}\int_{x_1}^{q_0} \! pdx 
\right) . 
\label{eqn:aboveb3}
\end{equation}
Since $p(x)$ is real everywhere, the choice of $x_1$ on the 
real axis is immaterial. 
The above formula determines the above the barrier reflection 
coefficient. 
It must be emphasised that to apply the above formula the 
exponent must be large so that $1-R$ is very nearly equal to unity. 

Finally consider a situation where the amplitudes of the 
reflection and incident wave are equal.  
The transmission coefficient is now an exponentially small 
quantity. 
This case corresponds to the standard tunnelling process. 
The problem is characterised by the presence of real turning 
points between which lies the classically forbidden region 
where the energy $E < V(x)$.  
For definiteness,  let there be two real turning points 
labelled $q_{-}$ and $q_{+}$. The potential $V(x)$, in the 
immediate vicinity of the turning points of the barrier, 
is assumed to be of the form
\begin{equation}
E- V(x) \approx F_0 (x-q_{\pm}), \qquad F_0 = -\left. 
{dV \over dx}\right|_{x = q_\pm}
\label{eqn:cond3}
\end{equation}
This assumption is equivalent to saying that the particle, 
near the turning points, moves in a homogeneous field.  
With this assumption, the amplitude of transmission $C_3$ 
is given by (refer to Ref.~\cite{landau3} page $181$)
\begin{equation}
C_3 =  \exp\left( -{1\over \hbar} \int_{q_{-}}^{q_{+}} \! 
|p(x)|dx  \right) 
\label{eqn:barrier1}
\end{equation} 
The transmission coefficient is then given by $|C_3|^2$. 
The above formula holds only when the exponent is large. 
In the derivation above, we have assumed that the 
semi-classical condition holds across the entire barrier 
except in the immediate vicinity of the turning points. 
In general, however, the semi-classical condition need not 
hold over the entire extent of the barrier. 
The potential, for example, could drop steeply enough so 
that Eqn.~(\ref{eqn:cond3}) is not valid. 
In these cases, the exact semi-classical equations have to 
be determined before applying the method of complex paths. 
The cases encountered in this paper all satisfy 
Eqn.~(\ref{eqn:cond3}). 

We now apply the above results to two potentials.  
The first is the well known inverted harmonic oscillator 
potential $V_1(x) = -g_1x^2$ with $g_1>0$ while the other 
is $V_2(x) = -g_2/x^2$ with $g_2 > 0$.  
The first potential arises when the propagation of a scalar 
field in a constant electric field background is studied.  
The second potential arises when the propagation of a scalar 
field in Schwarzchild-like spacetimes is considered in the 
vicinity of the horizon. 
 
\subsubsection{\it Application to the potential $V_1(x) = -g_1 x^2$} 
\label{subsubsec:potential1}
Consider the potential given by 
\begin{equation}
V_1(x) = -g_1 x^2 \label{eqn:pot1}
\end{equation}
where $g_1>0$ is a constant.  
This potential is the inverted harmonic oscillator potential 
and is discussed extensively in many places (see for 
example \cite{bandd82}, \cite{landau3}). 
We will follow the semi-classical treatment given in 
Ref.~\cite{landau3} and briefly review the calculation of 
the reflection and transmission coefficients for both the 
tunneling and over the barrier reflection cases.

The semi-classicality condition~(\ref{eqn:cond2}) for 
the above potential is, 
\begin{equation}
\left| {\hbar g_1 x \over \sqrt{2m} [E_1 + g_1 x^2]^{3/2}} 
\right| \ll 1 
\label{eqn:pot1cond}
\end{equation}
where $m$ is the mass of the particle and $E_1$ is its energy.  
The above condition definitely holds for large enough $|x|$ and 
for any value of $E_1$, either positive or negative. 
Therefore the motion  of a particle moving under such 
a potential is semi-classical for large enough $|x|$ and 
hence holds as $x \to \pm \infty$.   

Since the motion is semi-classical for large $|x|$, we can 
expand the momentum $p(x)$ as,
\begin{equation}
p(x) = \sqrt{2m\left( E_1 + g_1x^2 \right) } \approx  
\sqrt{2mg_1}\left( x + {E_1 \over 2g_1 x} \right) 
\label{eqn:electpot1} 
\end{equation}
Using Eqn.~(\ref{eqn:electpot1}), the  semi-classical wave 
functions can be written as follows with the following 
boundary conditions. 
As $x\to \infty$, we assume that the wave function is a 
right moving travelling wave $\psi_R$ while as $x\to -\infty$, 
it is a superposition of an incident wave of unit amplitude 
and a reflected wave given by $\psi_L$.  
Therefore, we have, 
\begin{eqnarray}
\psi_R = C_3 \xi^{i\varepsilon_1 - {1\over 2}} e^{i\xi^2/2} 
\qquad \qquad \qquad \qquad \qquad \; && 
\qquad \qquad (\xi \to +\infty) \label{eqn:electpot2a}  \\
\psi_L =  (-\xi)^{- i\varepsilon_1 - 
{1\over 2}} e^{- i\xi^2/2} + 
C_2 (-\xi)^{i\varepsilon_1 - {1\over 2}} e^{i\xi^2/2} && 
\qquad \qquad (\xi \to -\infty) 
\label{eqn:electpot2b}
\end{eqnarray}
where we have made the definitions
\begin{equation}
\xi = \left({2mg_1 \over \hbar^2}\right)^{\! 1/4} x \qquad 
; \qquad \varepsilon_1 =  {1\over \hbar}\sqrt{m \over 2g_1} E_1 
\label{eqn:electpot3}
\end{equation}
Following the variation of Eqn.~(\ref{eqn:electpot2a}) around 
a semi-circle of large radius $\rho$ in the {\it upper} 
half plane of the now complex variable $\xi$, we obtain 
\begin{equation}
C_2 = -iC_3 \exp\left( -\pi\varepsilon_1 \right) 
\label{eqn:electpot4}
\end{equation}
The conservation of particles is expressed by the condition 
that, 
\begin{equation}
|C_3|^2 + |C_2|^2 = 1 \label{eqn:pconservation}
\end{equation}
From Eqn.~(\ref{eqn:electpot4}) and 
Eqn.~(\ref{eqn:pconservation}), the transmission coefficient 
is, 
\begin{equation}
T = |C_3|^2 = {1 \over 1 + e^{-2\pi\varepsilon_1} } = 
{1 \over 1 + e^{-{1 \over \hbar} \pi \sqrt{2m/g_1}E_1} } 
\label{eqn:electpot5}
\end{equation}
while the reflection coefficient is 
\begin{equation}
R = |C_2|^2 =  {e^{-{1 \over \hbar} \pi \sqrt{2m/g_1}E_1} 
\over 1 + e^{-{1 \over \hbar} \pi \sqrt{2m/g_1}E_1} } 
\label{eqn:electpot6}
\end{equation}
Note that the passage through the {\it lower} half 
complex plane to determine $C_2$ is unsuitable since on the 
part of the path $\, -\pi < \phi< -\pi/2 \,$, where $\phi$ 
is the phase of the complex variable $\xi$, the incident wave 
(first term in Eqn.~(\ref{eqn:electpot2b})) is exponentially 
large compared with the reflected wave. 
The above formula holds for all energies $E_1$.  
This is because, even for negative energies, the 
semi-classical wave functions given in Eqns.~(\ref{eqn:electpot2a},
\ref{eqn:electpot2b}) are 
exactly the same with the boundary conditions being fully 
satisfied.   

If $E_1$ is large and negative, Eqn.~(\ref{eqn:electpot5}) 
gives $T \approx e^{-\pi\sqrt{2m/g_1}|E_1|/ \hbar}$ and 
thus $R\sim 1$. 
This is in accordance with the formula in Eqn.~(\ref{eqn:barrier1}). 
To apply Eqn.~(\ref{eqn:barrier1}) it is necessary to calculate 
the turning points first. 
The real turning points are $q_0 = -\sqrt{|E_1|/g_1}$ and 
$q_1 = \sqrt{|E_1|/g_1}$.  
Therefore,
\begin{eqnarray}
C_3 &=&  \exp\left( -{1\over \hbar} 
\int_{q_0}^{q_1} \! |p(x)|dx  \right) \nonumber \\
&=&  \exp\left(-{1\over \hbar}\sqrt{2mg_1}\int_{q_0}^{q_1} \! 
\left|\sqrt{x^2 - {|E_1|\over g_1}}\right| dx \right) 
\nonumber \\
&=& \exp\left(-{1\over 2\hbar}\pi \sqrt{2m/g_1}|E_1| \right) 
\label{eqn:electpot61}
\end{eqnarray} 
This gives the same answer. 

We can calculate the over the barrier reflection coefficient 
using Eqn.~(\ref{eqn:aboveb3}) for $E_1$ large and {\it positive}.  
The turning points now are given by solving the equation 
$p(q_0) = 0$ with the condition that $E_1>V_1(x)$ always.  
Since $E_1 > 0$, the turning points are  $q_0 = \pm i\sqrt{E_1/g_1}$. 
Choosing the positive sign for $q_0$ and setting $x_1 = 0$, 
the integral in Eqn.~(\ref{eqn:aboveb3}) is evaluated as follows:  
\begin{eqnarray}
\int_{0}^{q_0}\! p(x) dx &=& \sqrt{2mg_1} 
\int_{0}^{q_0}\!\sqrt{E_1/g_1 + x^2} \nonumber \\
&=& i\sqrt{2m/g_1}E_1\int_{0}^{1}\!\sqrt{1 - y^2} = 
{1 \over 4}i\pi \sqrt{2m/g_1}E_1
\label{eqn:electpot7}
\end{eqnarray}
Therefore, 
\begin{equation}
R = \exp\left( - {1 \over \hbar}\pi \sqrt{2m/g_1}E_1 \right) 
\label{eqn:electpot8} 
\end{equation}
The above formula can also be obtained directly from 
Eqn.~(\ref{eqn:electpot6}) by neglecting the exponential 
term compared to unity which means that the energy has 
to be large.  

\subsubsection{\it Application to the potential $V_2(x) 
= -g_2/x^2$} 
\label{subsubsec:potential2}
Consider the potential given by 
\begin{equation}
V_1(x) = -{g_2 \over x^2} \label{eqn:pot2}
\end{equation}
where $g_2$ is a positive constant. 
The potential has a singularity at the origin.  
This potential arises when the effective Schrodinger 
equation is calculated for Schwarzchild-like spacetimes. 
This aspect will be dealt with in later sections.  

The semi-classical condition~(\ref{eqn:cond2}) for this 
potential takes the form
\begin{equation}
\left|{\hbar g_2 \over \sqrt{2m}}
{1 \over [E_2 x^2 + g_2]^{3/2} }\right| \ll 1 
\label{eqn:pot2cond}
\end{equation} 
where $E_2$ is the energy. 
It is clear that the above relation holds for large $|x|$. 
It also holds for small $|x|$ if $\sqrt{2 m g_2} \gg \hbar$. 
Notice that the quasi-classicality condition for small $|x|$ 
is independent of the sign and magnitude of the energy $E_2$. 
For this potential, we will be concerned only with the small 
$|x|$ behaviour in contrast with the potential $V_1$.  
Since the motion is semi-classical for small $|x|$, 
we expand the momentum $p(x)$ as
\begin{equation}
p(x) = \sqrt{ 2m \left(E_2 + {g_2\over x^2} \right) } 
\approx {\sqrt{2mg_2} \over x} + \sqrt{m \over 2g_2}E_2 x 
\label{eqn:bhpot1}
\end{equation}
Notice the similarity between Eqn.~(\ref{eqn:electpot1}) 
and Eqn.~(\ref{eqn:bhpot1}). 

We will calculate the over the barrier reflection 
coefficient with $E_2 > 0$ and small, which will be 
of interest later. Using the expansion in 
Eqn.~(\ref{eqn:bhpot1}), the semi-classical wave 
functions with the following boundary conditions, 
namely, that for $x>0$ the wave function is a right 
moving travelling wave while it is a superposition of 
an incident wave of unit amplitude and reflected wave 
for $x<0$,  are
\begin{eqnarray}
\psi_R = C_3 \xi^{ i\varepsilon_2 + {1\over 2}} 
e^{ i\xi^2/2} \qquad \qquad \qquad && \qquad \qquad 
(\xi > 0) \label{eqn:bhpot2a} \\ 
\psi_L =  (-\xi)^{- i\varepsilon_2 + {1\over 2}} 
e^{- i\xi^2/2} + C_2 (-\xi)^{ i\varepsilon_2 + {1\over 2}} 
e^{i\xi^2/2} && \qquad \qquad (\xi < 0) 
\label{eqn:bhpot2b}
\end{eqnarray}
where we have made the definitions
\begin{equation}
\xi = \left({mE_2^2 \over 2g_2}\right)^{1/4}\! x \qquad 
; \qquad \varepsilon_2 = {\sqrt{2mg_2} \over \hbar} 
\label{eqn:bhpot3}
\end{equation}
Following the variation of Eqn.~(\ref{eqn:bhpot2a}) around 
an small semi-circle of radius $\rho < \sqrt{g_2/|E_2|}$ 
(in contrast to the potential $V_1$ where the radius 
$\rho$ was large) in the {\it upper } half complex plane, 
we obtain,
\begin{equation}
 C_2 = C_3 \exp\left( -\pi\varepsilon_2 +{i\pi \over 2} 
\right) 
\label{eqn:bhpot4}
\end{equation}
Setting $T = |C_3|^2 = 1$ and $R = |C_2|^2$, we 
finally obtain, 
\begin{equation}
R = T e^{-2\pi\varepsilon_2} = T 
e^{-{1 \over \hbar}2\pi\sqrt{2mg_2}} 
\label{eqn:bhpot5}
\end{equation}
Using the normalisation condition $R + T = 1$, we 
obtain,
\begin{equation}
T = { 1 \over 1 + e^{-{1 \over \hbar}2 \pi\sqrt{2mg_2}} } 
\quad {\rm  and } \quad R =  { e^{-{1\over \hbar}2 \pi\sqrt{2mg_2}} 
\over 1 + e^{-{ 1\over \hbar}2 \pi\sqrt{2mg_2}} } 
\label{eqn:bhpot51}
\end{equation}
Notice that the above result is independent of the energy 
$E_2$ and hence holds for $E_2 < 0$ too. For small $|x|$, 
the lack of dependence on $E_2$ is not too surprising since 
the contour is such that it is not too close to the real 
turning points $q_0 = \pm \sqrt{g_2/|E_2|}$.  
Anyway, when $E_2\sim 0^{+}$, $\rho$ is ``large'' and 
therefore the contour is chosen to lie in the upper complex 
plane for the same reason as given in the analysis of the 
potential $V_1$ in the previous section. 

Now, we will derive the result in Eqn.~(\ref{eqn:bhpot5}) 
using Eqn.~(\ref{eqn:aboveb2}).  
The complex turning points $q_0$ are the roots of the equation 
$E_2 = -g_2/q_0^2$ where $E_2>0$ and therefore, the turning 
points are $q_0 = \pm i\sqrt{g_2/E_2} = \pm i p_0$.  
Hence, we have to evaluate the integral,
\begin{equation}
\int_C \! pdx = \sqrt{2mE_2}\int_C \! 
\sqrt{1 + {p_0^2 \over x^2}} dx 
\label{eqn:bhpot7}
\end{equation}
where the contour $C$ encircles the point $x = ip_0$ in 
the upper half complex plane.  
However, since there is a singularity at the origin, 
we cannot deform the contour as was done when deriving 
Eqn.~(\ref{eqn:aboveb3}).  
Therefore, as a means of regularisation, we modify the 
potential to 
\begin{equation}
V_{\rm mod}(x) = -{g_2 \over x^2 + \epsilon^2} 
\label{eqn:bhpot8}
\end{equation}
where the limit $\epsilon \to 0$ must be taken at the end 
of the calculation.   
The turning points for the modified potential are 
$x_{\rm mod} = \pm i\sqrt{\epsilon^2 + g_2/E_2}$ while the 
poles of the modified potential are at 
$x = \pm i\epsilon < x_{\rm mod}$. 
Even in this case, there is a singularity on the path of 
integration which contributes to the integral rather than 
the turning point.  
Therefore, integrating upto $+i\epsilon$ using the modified 
potential and back, we obtain,  
\begin{eqnarray}
\int_C \! pdx &=& \lim_{\epsilon \to 0} 
2\sqrt{2mE_2}\int_{0}^{i\epsilon}\! 
\sqrt{ 1 + {p_0^2 \over x^2 + \epsilon^2}} dx \nonumber \\
&=& \lim_{\epsilon \to 0} 2i\sqrt{2mE_2}\epsilon \int_{0}^{1} 
\! dy \sqrt{ 1 +  {p_0^2/\epsilon^2 \over 1 - y^2} } 
\nonumber \\
&\approx & 2i\sqrt{2mE_2}p_0 \int_o^1 \! {dy \over \sqrt{1-y^2}} 
\nonumber \\
&=& i\pi \sqrt{2mE_2}p_0 = i\pi \sqrt{2mg_2} 
\label{eqn:bhpot9}
\end{eqnarray}   
We therefore recover the result given in Eqn.~(\ref{eqn:bhpot5}). 
From the above calculation it is clear that, due to the 
singularity at the origin, the reflection coefficient has no 
contribution from the turning point at all.

\section{Particle production in an uniform Electric field} 
\label{sec:electric}

We will now study a system consisting of a minimally coupled 
scalar field $\Phi$ propagating in flat spacetime in an 
uniform electric field background. 
We consider two gauges, one a time dependent gauge while 
the other is a space dependent gauge and show how the 
tunnelling interpretation can be used to obtain the 
standard gauge invariant result in each case.  
We will not derive the standard result here.  

\subsection{Time dependent gauge} 
\label{subsec:electrictime}

The four vector potential $A^{\mu}$ giving rise to a 
constant electric field in the $x$ direction is assumed 
to be of the form
\begin{equation}
A^{\mu} = (0, -E_0t, 0,0) \label{eqn:gauge1}
\end{equation}
The electric field is ${\bf E} = E_0\hat{\bf x}$.  
The minimally coupled scalar field $\Phi$ propagating in 
flat spacetime, satisfies the Klein-Gordon equation,
\begin{equation}
\left( (\partial_{\mu} + iqA_{\mu})(\partial^{\mu} + iqA^{\mu}) 
+ m^2\right)\Phi = 0 
\label{eqn:electric11}
\end{equation}
where $m$ is the mass and $q$ is the charge of the field. 
The mode functions of $\Phi$ can be expressed in the form 
$\Phi(t,{\bf x}) = f_{\bf k}(t)e^{i{\bf k}\cdot {\bf x}}$ 
where $f_{\bf k}(t)$ satisfies the equation,
\begin{equation}
{d^2 \over dt^2}f_{\bf k} + \left[ m^2 + k_{\perp}^2 + 
(k_x + qE_0t)^2\right]f_{\bf k} = 0 \; ;\quad {\bf k}_{\perp} 
= (k_y,k_z) .
\label{eqn:electric12}
\end{equation}
Introducing the variables,
\begin{equation}
\tau  = \sqrt{qE_0}t + \left(\omega/\sqrt{qE_0}\right) \; 
; \;  \lambda = (k_{\perp}^2 + m^2)/qE_0 
\label{eqn:electric13}
\end{equation}
we obtain the equation,
\begin{equation}
-{d^2 \over d\tau^2}f_{\bf k} - \tau^2 f_{\bf k} 
= \lambda f_{\bf k} 
\label{eqn:electric14} 
\end{equation}
The above equation is essentially a Schrodinger equation in an 
inverted oscillator potential with a positive ``energy'' $\lambda$. 
Since the energy is positive, the problem is essentially an 
over the barrier reflection problem.  Using the results of 
section~(\ref{subsubsec:potential1}), we can calculate the 
reflection and transmission  coefficients exactly as
\begin{equation}
R =  {e^{-\pi\lambda} \over 1 + e^{-\pi\lambda} } \quad ; 
\quad T = {1 \over 1 + e^{- \pi\lambda} } 
\label{eqn:electric15}
\end{equation}
where we have put $\hbar = 2m = \omega_0/2 = 1$ and set 
$a^2 = \lambda$ in Eqns.~(\ref{eqn:electpot5}, 
\ref{eqn:electpot6}). 
To identify the Bogoliubov coefficients $\alpha_{\lambda}$ 
and $\beta_{\lambda}$, we recast the normalisation condition 
$R+T=1$ in the form,
\begin{equation}
{1 \over T} - {R \over T} = 1
\end{equation}
and then identify $|\beta_\lambda|^2$ with $R/T$ and 
$|\alpha_\lambda|^2$ with $1/T$. 
Therefore, the Bogoliubov coefficients are given by, 
\begin{eqnarray}
|\beta_\lambda|^2 &=& e^{-\pi\lambda} = 
e^{-\pi(k_{\perp}^2 + m^2)/qE_0} \nonumber \\
|\alpha_\lambda|^2 &=&   e^{-\pi\lambda} + 1 = 
e^{-\pi(k_{\perp}^2 + m^2)/qE_0} + 1 
.\label{eqn:electric16}
\end{eqnarray}
Now, the transmission and reflection coefficients are time 
reversal invariant.  
They are dependent only on the energy (magnitude and sign) 
and for symmetric potentials, independent of the direction 
in which the boundary conditions are applied. 
To obtain a dynamical picture of particle production, 
we have to interpret these quantities suitably. 
In the present case, the following interpretation seems adequate. 
A purely positive frequency wave with amplitude square $T$ 
in the infinite past, $t \to -\infty$, evolves into a 
combination of positive and negative frequency waves in 
the infinite future $t \to \infty$ with the negative frequency 
waves having an amplitude square $R$ and the positive frequency 
waves having an amplitude unity. 
The quantity  $R/T$  determines the overlap between the 
negative frequency modes in the distant future and the 
positive frequency modes in the distant past (the notation 
here differs from the treatment given in \cite{paddy91}, 
\cite{brout95}). 
This is identified with the modulus square of the 
Bogoliubov coefficient $\beta_{\lambda}$ which is the 
particle production per mode $\lambda$. 
Using the normalisation condition satisfied by the Bogoliubov 
coefficients, $|\alpha_\lambda|^2 - |\beta_{\lambda}|^2 = 1$,
$|\alpha_\lambda|^2$ can be calculated to be $1/T$.
Once the Bogoliubov coefficients have been identified, the 
effective lagrangian can be easily calculated.  
This derivation will not be repeated here.  
We refer the reader to Ref.~\cite{brout95} and 
Ref.~\cite{paddy91} for the explicit calculation.

Note that the particular interpretation given in this 
case is due to its similarity with the more rigourous 
calculation by quantum field theoretic methods. 
In the next section, where we discuss the space dependent 
gauge, we will be forced to adopt a different interpretation 
in order to identify particle production.

\subsection{Space dependent gauge} 
\label{subsec:electricspace}
The four vector potential $A^{\mu}$ giving rise to a constant 
electric field in the $x$ direction is now assumed to be 
of the form
\begin{equation}
A^{\mu} = (-E_0x, 0,0,0) \label{eqn:gauge2}
\end{equation}
The electric field is ${\bf E} = E_0\hat{\bf x}$.  
The field  $\Phi$ satisfies Eqn.~(\ref{eqn:electric11}) as 
before.
Substituting for the potential $A^{\mu}$ from 
Eqn.~(\ref{eqn:gauge2}) into Eqn.~(\ref{eqn:electric11}), 
we obtain,
\begin{equation}
\left( \partial_t^2 - \nabla^2 - 2iqE_0x\partial_t - 
q^2E_0^2x^2 + m^2\right)\Phi = 0 
\label{eqn:electric22}
\end{equation}
We write $\Phi$ in the form
\begin{equation}
\Phi = e^{-i\omega t} e^{i k_y y + ik_zz} \phi(x) 
\label{eqn:electric23}
\end{equation}
and obtain the differential equation satisfied 
by $\phi$ as
\begin{equation}
{d^2 \phi \over dx^2} + \left( (\omega + qE_0x)^2 - 
k_{\perp}^2 - m^2 \right)\phi = 0
\label{eqn:electric24}
\end{equation}
where we have used the notation $k_{\perp}^2 = k_y^2 + k_z^2$. 
Making  the following change of variables 
\begin{equation}
\rho  = \sqrt{qE_0}x + \left(\omega/\sqrt{qE_0}\right) 
\quad ; \quad
\lambda = (k_{\perp}^2 + m^2)/qE_0
\label{eqn:electric25}
\end{equation}
into the differential equation for $\phi$, it reduces 
to the form,
\begin{equation}
-{d^2 \phi \over d\rho^2} - \rho^2\phi = -\lambda\phi 
\label{eqn:electric26}
\end{equation}
In this form, we see that the above differential equation 
has the form of an effective Schrodinger equation with 
an inverted harmonic oscillator potential and an 
negative energy $-\lambda$.  
If we apply the results of section~(\ref{subsubsec:potential1}), 
we obtain the result for tunneling through the barrier.  
Following the treatment in Ref.~\cite{paddy91} and using 
the results of section~(\ref{subsubsec:potential1}) we can 
calculate the reflection and transmission  coefficients exactly 
as
\begin{equation}
R =  {e^{ \pi\lambda} \over 1 + e^{\pi\lambda} }\quad ; 
\quad T = {1 \over 1 + e^{\pi\lambda} } 
\label{eqn:electric27}
\end{equation}
where we have put $\hbar = 2m = \omega_0/2 = 1$ and set 
$a^2 = \lambda$ in Eqns.~(\ref{eqn:electpot5}, \ref{eqn:electpot6}). 
We cast the renormalisation condition $R+T=1$ in the form
\begin{equation}
{1 \over R} - {T\over R} = 1
\end{equation}
and then identify the rate of particle production per mode with 
$T/R$. 
The interpretation of particle production using the tunnelling 
picture now proceeds as follows. 
A right moving travelling wave of amplitude square $1/R$ is 
incident on the potential. 
A fraction $T/R$ is transmitted through it and a wave of unit 
amplitude is scattered back.  
The tunnelling probability, which is $T/R$, is interpreted as 
the rate at which particles are being produced by the 
background electric field. 
This matches exactly with the expression for 
$|\beta_{\lambda}|^2$ given in Eqn.~(\ref{eqn:electric16}). 
With this interpretation, we recover the usual gauge 
independent result.

In summary, it is seen that by a judicious choice of interpretation 
of the transmission and reflection coefficients in each of the 
two gauges, the standard gauge invariant result can be obtained. 
In fact, the tunnelling interpretation can also be used for 
gauges of the form
\begin{equation}
A^{\mu} = (-E_0x \pm E_1t, 0,0,0) \quad ; \quad A^{\mu} = 
(0,-E_0t \pm E_1x,0,0)
\end{equation}
where the condition $|E_1| \neq |E_0|$ holds strictly. 
Depending on the magnitude of $E_1$, the problem reduces 
to either an ``over the barrier reflection''  
or a ``tunnelling through the barrier'' process.  
The transmission and reflection coefficients do not depend 
on $E_1$ as it ought to be.  
But the tunnelling interpretation is not always valid.  
For instance, setting $|E_1| = |E_0|$ in the above gauges 
reduces the problem to a first order differential equation 
instead of a second order one. 
(This case will be considered later in a future publication.) 
Further, one can show that even in a time independent magnetic 
field, solving the effective Schrodinger equation gives 
non-zero transmission and reflection coefficients~\cite{sriram97}.  
Since these coefficients are non-zero, applying the tunnelling 
picture implies particle production which is contrary to the 
result obtained by Schwinger in~\cite{schwinger}. 
Hence, the tunnelling picture does not always produce consistent 
results.  
In the next section, we look at Schwarzchild-like spacetimes 
with a horizon.  
Here too, we construct the tunnelling picture near the horizon 
in order to explain particle production. 
We shall see that such a picture does not produce the correct 
results.

\section{Particle Production in spacetimes with  Horizon} 
\label{sec:schwarzchild}

Hawking's result that a black hole radiates is essentially a 
semi-classical result.  
The thermal radiation results because of the presence of a 
horizon in the spacetime structure. 
We will review briefly the conventional derivation of the 
thermal radiation using path integrals.  
Consider a patch of spacetime, which in a suitable co-ordinate 
system, has one of the following forms (we assume $c=1$):
\begin{equation}
ds^2 = B(r)dt^2 - B^{-1}(r) dr^2 - r^2(d\theta^2 + 
\sin^2(\theta)d\phi^2) 
\label{eqn:metric2}
\end{equation}
or
\begin{equation}
ds^2 = B(x)dt^2 - B^{-1}(x) dx^2 - dy^2-dz^2 
\label{eqn:metric3}
\end{equation}
where $B(r)$ and $B(x)$ are functions of $r$ and $x$ respectively.  
The horizon in the above spacetimes is indicated by the surface 
$r=r_0$ ($x=x_0$) at which $B(r)$ ($B(x)$) vanishes.  
We further assume that $B'(r)= dB/dr$ ($B'(x)=dB/dx$) is finite 
and non-zero at the horizon.  
Co-ordinate systems of the form (\ref{eqn:metric2}) can be 
introduced in parts of the Schwarzchild and de-Sitter spacetimes 
while that of the form (\ref{eqn:metric3}) with the choice 
$B(x) = 1 + 2gx$ represents a Rindler frame in flat spacetime.  
Given the co-ordinate system of (\ref{eqn:metric2}) say, in 
some region {\cal R}, we first verify that there is no physical 
singularity at the horizon, which in the case of the 
Schwarzchild black hole, is at the co-ordinate value $r_0=2M$ 
where $M$ is the mass of the black hole.  
Having done that, we extend the geodesics into the past and 
future and arrive at two further regions of the manifold not 
originally covered by the co-ordinate system in (\ref{eqn:metric2}) 
(the Kruskal extension).  
It is now possible to show that the probability for a particle 
with energy $E$ to be lost from the region {\cal R} in relation 
to the probability for a particle with energy $E$ to be gained 
by the region {\cal R} is given by the relation
\begin{equation}
P_{\rm loss} = P_{\rm gain} e^{-\beta E} 
\label{eqn:thermal}
\end{equation}
where $\beta = 8 \pi M$.  This is equivalent to assuming that 
the region {\cal R} is bathed in radiation at temperature 
$\beta^{-1}$.  
In the derivation given in the paper by Hartle and Hawking 
\cite{hawking76}, thermal radiation is derived using the 
semiclassical kernel by an analytic continuation in the 
time co-ordinate $t$ to complex values and it was shown that 
the probability of emission (loss) from the past horizon was 
related to absorption (gain) into the future horizon by the
relation~(\ref{eqn:thermal}).

Since all the physics is contained in the $(t,r)$ plane 
(or the $(t,x)$ plane), we will discuss Hawking radiation 
in $1+1$ dimensions in the following sections and later show 
in the appendix that the  results generalise naturally to 
$3 + 1$ dimensions.
We first derive the semi-classical result in the $(r, t)$ 
(or $(x,t)$) plane by applying a certain prescription to bypass 
the singularity  encountered at the horizon.  
After this, we reduce the problem of the Klein-Gordon field 
propagating in the Schwarzchild spacetime to an  effective 
Schrodinger problem in (1+1) dimensions and rederive the 
semi-classical result by using the same prescription on the 
semi-classical wave functions.
Then the results of section~(\ref{sec:facets}) are used to 
compute the ratio of the transmission and reflection coefficients 
and we show that this ratio satisfies an equation of the form 
\begin{equation}
R = T e^{-{1 \over 2}\beta E}
\end{equation}
where $\beta^{-1} = (8\pi M)^{-1}$ is the expected Hawking 
temperature. 
This can be interpreted to mean thermal radiation in exactly 
the same way as Eqn.~(\ref{eqn:thermal}) but with a temperature 
{\it twice} the expected value. 
This proves that the tunnelling picture, when naively applied, 
does not give the same result as the semi-classical analysis does.  
A possible qualitative explanation of this result will also be given.

\subsection{Hawking Radiation in 1+1 dimensions}
\label{sec-hawking}
Consider a certain patch of spacetime in (1+1) dimensions 
which in a suitable co-ordinate system has the line element, 
(with $c=1$) 
\begin{equation}
ds^2 = +B(r)dt^2 - B^{-1}(r) dr^2  \label{eqn:metric1}
\end{equation}
where $B(r)$ is an arbitrary function of $r$.   
We assume that the function $B(r)$ vanishes at some $r = r_0$ 
with $B'(r)=dB/dr$ being finite and nonzero at $r_0$. 
The point $r=r_0$ indicates the presence of a horizon.  
It can be easily verified that no physical singularity exists 
at the horizon since the only component of the curvature 
tensor ${\cal R}_{trtr} = -(1/2) (d^2B(r)/dr^2)$ does not 
become infinite at the horizon. 
Therefore, near the horizon, we may expand $B(r)$ as
\begin{equation}
B(r) = B'(r_0)(r - r_0) + {\cal O}[(r-r_0)^2] = R(r_0)(r-r_0).
\label{eqn:horizon}
\end{equation}

We now consider a minimally coupled scalar field $\Phi$ with 
mass $m_0$ propagating in the spacetime represented by the 
metric (\ref{eqn:metric1}).  
The equation satisfied by the scalar field is, 
\begin{equation}
\left(\Box + {m^2_0 \over \hbar^2}\right)\Phi
= 0 \label{eqn:kg}
\end{equation}
where the $\Box$ operator is to be evaluated using metric 
(\ref{eqn:metric1}).  
Expanding the LHS of equation~(\ref{eqn:kg}), 
one obtains, 
\begin{equation}
{1 \over B(r)} {\partial^2 \Phi \over \partial t^2} -
{\partial \over \partial r} \left( B(r) {\partial \Phi 
\over \partial r} \right) = -{m_0^2 \over \hbar^2} \Phi .
\label{eqn:kgr}
\end{equation}
The semiclassical wave functions satisfying the above equation 
are obtained by making the standard ansatz for $\Phi$ which is,
\begin{equation}
\Phi(r,t) = \exp\left[ {i \over \hbar} S(r,t)\right] 
\label{eqn:semi}
\end{equation}
where $S$ is a function which will be expanded in powers of $\hbar$.   
Substituting into the wave equation~(\ref{eqn:kgr}), we obtain,
\begin{eqnarray}
\left[ {1 \over B(r)} \left( {\partial S \over \partial t} 
\right)^2 - B(r)\left( {\partial S \over \partial r} 
\right)^2 -m^2_0 \right] && +  \nonumber \\
&& \left({\hbar \over i}\right) \left[ {1 \over B(r)} 
{\partial^2 S \over \partial t^2}- B(r) {\partial^2 S \over 
\partial r^2} -  { dB(r) \over dr}{\partial S \over \partial r} 
\right] = 0 
\label{eqn:seqn1}
\end{eqnarray}
Expanding $S$ in a power series of $(\hbar/i)$,
\begin{equation}
S(r,t) = S_0(r,t) + \left({\hbar \over i}\right) S_1(r,t) +
 \left({\hbar \over i}\right)^2 S_2(r,t) \ldots 
\label{eqn:exp}
\end{equation}
and substituting into Eqn~(\ref{eqn:seqn1}) and neglecting 
terms of order $(\hbar/i)$ and greater, we find to the lowest 
order,
\begin{equation}
{1 \over B(r)} \left( {\partial S_0 \over \partial t} 
\right)^2 - B(r)\left( {\partial S_0 \over \partial r} 
\right)^2 -m_0^2 = 0 
\label{eqn:seqn2}
\end{equation}
Eqn~(\ref{eqn:seqn2}) is just the Hamilton-Jacobi equation 
satisfied by a particle of mass $m_0$ moving in the spacetime 
determined by the metric~(\ref{eqn:metric1}).  
The solution to the above equation is 
\begin{equation}
S_0(r,t) = -Et \pm \int^r {dr \over B(r)} 
\sqrt{E^2 - m^2_0 B(r)} \label{eqn:ssol1}
\end{equation}
where $E$ is a constant and is identified with the energy. 
Notice that in the case of $m_0=0$, Eqn~(\ref{eqn:seqn2}) can 
be exactly solved with the solution
\begin{equation}
S_{m_0=0}(r,t) = F_1(t - r*) + F_2(t + r*) 
\label{eqn:ssol2}
\end{equation}
where  the ``tortoise'' co-ordinate $r*$ is defined by 
\begin{equation}
r* = \int {dr \over B(r)},
\end{equation}
and $F_1$ and $F_2$ are arbitrary functions.  
If $F_1$ is chosen to be $F_1 = -Et + Er*$ and $F_2$ chosen 
to be $F_2 = -Et - Er*$, then it is clear that the solution 
given in Eqn~(\ref{eqn:ssol2}) is the same as that in 
Eqn~(\ref{eqn:ssol1}) with $m_0$ set to zero.  
Therefore, in the case $m_0=0$, the semiclassical ansatz is exact. 
In the following analysis we will specialise to the case 
$m_0=0$ for simplicity.  The case $m_0\neq 0$ will be 
considered later.  
The essential results do not change in any way.  

Knowing the semiclassical wave function, the semiclassical 
kernel \, $K(r_2,t_2; r_1,t_1)$ for the particle to propagate 
from $(t_1,r_1)$ to $(t_2,r_2)$ in the saddle point 
approximation can be written down immediately.
\begin{equation}
K(r_2,t_2; r_1,t_1) = N \exp\left( {i\over \hbar} 
S(r_2,t_2; r_1,t_1) \right)
\end{equation} 
where $S$ is the action functional satisfying the classical 
Hamilton-Jacobi equation in the 
massless limit and $N$ is a suitable normalization constant.  
$S(r_2,t_2; r_1,t_1)$ is given by the relation
\begin{equation}
S(r_2,t_2; r_1,t_1)= S(2,1) = -E(t_2-t_1) \pm E 
\int^{r_2}_{r_1} {dr \over B(r)}.  
\label{eqn:ssol3}
\end{equation}
The sign ambiguity (of the square root) is related to 
the ``outgoing'' ($\partial S /\partial r\,>0$) or 
``ingoing'' ($\partial S /\partial r\,<0$) nature of the 
particle.
As long as points $1$ and $2$, between which the transition 
amplitude is calculated, are on the same side of the horizon 
(i.e. both are in the region $r>r_0$ or in the region $r<r_0$), 
the integral in the action is well defined and real. 
But if the points are located on opposite sides of the horizon 
then the integral does not exist due to the divergence of 
$B^{-1}(r)$ at $r=r_0$.

Therefore, in order to obtain the probability amplitude for 
crossing the horizon we have to give an extra prescription 
for evaluating the integral \cite{paddy91}.  
Since the point $B=0$ is null we may carry out the calculation 
in Euclidean space or ---equivalently---use an appropriate 
$i\epsilon$ prescription to specify the complex contour over 
which the integral has to be performed around $r=r_0$.  
The prescription we use is that we should take the contour 
for defining the integral to be an infinitesimal semicircle 
{\it above} the pole at $r=r_0$ for outgoing particles on 
the left of the horizon and ingoing particles on the right.  
Similarly, for ingoing particles on the left and outgoing 
particles on the right of the horizon (which corresponds to 
a time reversed situation of the previous cases) the contour 
should be an an infinitismal semicircle {\it below} the 
pole at $r=r_0$. 
Equivalently, this amounts to pushing the singularity at 
$r=r_0$ to $r = r_0 \mp i\epsilon$ where the upper sign 
should be chosen for outgoing particles on the left and 
ingoing particles on the right while the lower sign should 
be chosen for ingoing particles on the left and outgoing 
particles on the right. 
For the Schwarzchild case, this amounts to adding an 
imaginary part to the mass since $r_0 = 2M$.  

Consider therefore, an outgoing particle 
($\partial S /\partial r\,>0$) at $r=r_1<r_0$. 
The modulus square of the amplitude for this particle to 
cross the horizon gives the probability of emission of 
the particle.  
The contribution to $S$ in the ranges $(r_1,r_0-\epsilon)$ 
and $(r_0+\epsilon,r_2)$ is real. 
Therefore, choosing the contour to lie in the upper 
complex plane,
\begin{eqnarray}
S[{\rm emission}] &=& -\lim_{\epsilon \to 0} 
\int^{r_0+\epsilon}_{r_0-\epsilon} {dr \over B(r)} 
+ \; ({\rm real \; part})  \nonumber \\
&=& {i \pi E \over R(r_0)}+ ({\rm real \; part})
\end{eqnarray}
where the minus sign in front of the integral corresponds 
to the initial condition that $\partial S /\partial r\,>0$ 
at $r=r_1<r_0$.  
For the sake of definiteness we have assumed $R(r_0)$ in 
Eqn~(\ref{eqn:horizon}) to be positive, so that 
$B(r)<0$ when $r<r_0$.  
(For the case when $R<0$, the answer has to be modified by a 
sign change.)  
The same result is obtained when an ingoing particle 
($\partial S /\partial r\,<0$) is considered at 
$r=r_1<r_0$.  
The contour for this case must be chosen to lie in the 
lower complex plane. 
The amplitude for this particle to  cross the horizon  is 
the same as that of the outgoing particle due to the time 
reversal invariance symmetry obeyed by the system.     

Consider next, an ingoing particle 
($\partial S /\partial r\,<0$) at $r=r_2>r_0$.  
The modulus square of the amplitude for this particle to cross 
the horizon gives the probability of absorption of the 
particle into the horizon. 
Choosing the contour to lie in the upper complex plane, 
we get, 
\begin{eqnarray}
S[{\rm absorption}] &=& -\lim_{\epsilon \to 0} 
\int^{r_0-\epsilon}_{r_0+\epsilon} {dr \over B(r)} + 
\; ({\rm real \; part})  \nonumber \\
&=& -{i \pi E \over R(r_0)} + ({\rm real \; part})
\end{eqnarray}
The same result is obtained when an outgoing particle 
($\partial S /\partial r\,>0$) is considered at $r=r_2>r_0$.  
The contour for this case should be in the lower complex 
plane and the amplitude for this particle to cross the 
horizon  is the same as that of the ingoing particle due 
to time reversal invariance.

Taking the modulus square to obtain the probability $P$, 
we get,
\begin{equation}
P[{\rm emission}] \propto \exp\left(-{2 \pi E \over R(r_0)} 
\right)
\end{equation}
and 
\begin{equation}
P[{\rm absorption}] \propto \exp\left({2 \pi E \over R(r_0)} 
\right)
\end{equation}
so that
\begin{equation}
P[{\rm emission}] = \exp\left(-{4 \pi E \over R(r_0)} 
\right)P[{\rm absorption}]. \label{eqn:ssys}
\end{equation}
Now time reversal invariance implies that the probability 
for the emission process is the same as that for the absorption 
process proceeding backwards in time and {\it vice versa}. 
Therefore we must interpret the above result as saying that the 
probability of emission of particles is not the same as the 
probability of absorption of particles. 
In other words, if the horizon emits  particles at some time 
with a certain emission probability, the probability of 
absorption of particles at the same time is different from 
the emission probability.  
This result shows that it is more likely for a particular 
region to gain particles than lose them.  
Further, the exponential dependence on the energy allows one to 
give a `thermal' interpretation to this result.  
In a system with a temperature $\beta^{-1}$ the absorption 
and emission probabilities are related by 
\begin{equation}
P[{\rm emission}]= \exp(-\beta E) P[{\rm absorption}] 
\label{eqn:thermalsys}
\end{equation}
Comparing Eqn~(\ref{eqn:thermalsys}) and Eqn~(\ref{eqn:ssys}), 
we identify the temperature of the horizon in terms of $R(r_0)$.
Eqn~(\ref{eqn:ssys}) is based on the assumption that $R>0$.  
If $R<0$ there will be a change of sign in the equation.  
Incorporating both the cases, the general formula for 
the horizon temperature is
\begin{equation}
\beta^{-1} = {|R| \over 4\pi} \label{eqn:stemp}
\end{equation}
For the Schwarzchild black hole,
\begin{equation}
B(r) = \left( 1 - {2M \over r}\right) \approx 
{1 \over 2M}(r-2M) + {\cal O}[(r-2M)^2]
\end{equation}
giving $R=(2M)^{-1}$, and the temperature $\beta^{-1}= 
1/8\pi M$.  
For the de-Sitter spacetime,
\begin{equation}
B(r) = (1-H^2r^2) \approx 2H(H^{-1}-r) = -2H(r-H^{-1})
\end{equation}
giving $\beta^{-1}= H/2\pi$.  
Similarly for the Rindler spacetime
\begin{equation}
B(r) = (1 + 2gr) = 2g(r + (2g)^{-1})
\end{equation}
giving $\beta^{-1}= g/2\pi$.  
The formula for the temperature can be used for more complicated 
metrics as well and gives the same results as obtained by 
more detailed methods.

The prescription given for handling the singularity is 
analogous to the analytic continuation in time proposed 
by Hawking~\cite{hawking76} to derive Black hole radiance.  
If one started out on the left of the horizon and went 
around the singularity $r=r_0$ by a $2\pi$ rotation instead 
of a rotation by $\pi$, it can be easily shown that it has 
the effect of taking the Kruskal co-ordinates 
$(v,u)$ to $(-v,-u)$.  
A full rotation by $2\pi$ around the singularity can be 
split up into two parts to give the amplitude for emission 
and subsequent absorption of a particle with energy $E$. 
Since the amplitudes for the two processes are not the same 
in the presence of a horizon, one obtains the usual 
Hawking radiation given in Eqn~(\ref{eqn:ssys}) with 
the  value of $R(r_0)$ being $(2M)^{-1}$. 
This process is similar to that given in \cite{hawking76} 
which relates the amplitudes involving the 
past and future horizons.  
In Hawking's paper, analytically continuing the time 
variable $t$ to $t-4Mi\pi$ takes the Kruskal co-ordinates  
$(v,u)$ to $(-v,-u)$ and since the path integral kernel 
is analytic in a strip of $4Mi\pi$ below the real $t$ axis, 
Hawking radiation is obtained by deforming the contour 
of integration appropriately.

When $m_0 \neq 0$, the validity of the semi-classical 
ansatz must be verified.  
To do this, consider the perturbative expansion~(\ref{eqn:exp}).  
Retaining the terms of order $\hbar/i$ and neglecting 
higher order terms, one finds, upon substituting for $S_0$ 
given by the relation~(\ref{eqn:ssol1}) and solving for $S_1$,
\begin{equation}
S_1 = -E_1t \pm EE_1 \int {dr \over B(r)}
{1\over \sqrt{E^2 - m_0^2 B(r)}} - 
{1 \over 4}\ln(E^2 - m_0^2 B(r)) 
\end{equation}
where $E_1$ is a constant.  From the above equation, it 
is seen that $S_1$ has a singularity of the same 
order as $S_0$ at $r=r_0$.  
When calculating the amplitude  to cross the horizon, 
the contribution from the singular term just appears as 
a phase factor multiplying the semiclassical kernel and 
is inconsequential. 
The non-singular finite terms do contribute to the kernel 
but they contribute the same amount to $S[{\rm emission}]$ 
and $S[{\rm absorption}]$ and they do not affect the relation 
between the probabilities $P[{\rm emission}]$ 
and $P[{\rm absorption}]$. 
Subsequent calculation of the terms $S_2$, $S_3$, and so on, 
show that all these terms have a singularity at the 
horizon of the same order as that of $S_0$. 
Their contribution to the probability amplitude is just 
a set of terms multiplied by powers of $\hbar$ 
which can be neglected.  
From this we may conclude that the semiclassical 
ansatz, in the perturbative limit, is a valid one.

\subsection{Reduction to an effective Schrodinger 
Problem in (1+1) dimensions} 
\label{sec:schrodinger}
Consider the relativistic equation for the wave 
function $\Phi$ in Eqn~(\ref{eqn:kgr}).  
We include the mass $m_0$ here but later we shall see that 
it does not appear in the final reduced Schrodinger equation.  
We first set $\Phi = e^{-iEt/\hbar}\Psi(r)$ to obtain, 
\begin{equation}
B(r) {d^2 \Psi \over dr^2} + {d B(r) \over dr} 
{d\Psi \over dr} + \left( {E^2 \over \hbar^2 B(r)} - 
{m_0^2 \over \hbar^2} \right) \Psi = 0 
\label{eqn:sch1}
\end{equation}
We then make the substitution
\begin{equation}
\Psi(r) = P(r) Q(r) \; \; {\rm with} \; P(r) = 
{1 \over \sqrt{B(r)}} 
\label{eqn:schsubs}
\end{equation}
to get the equation,
\begin{equation}
 -{d^2 Q(r) \over dr^2} - \left[ -{B''(r) \over 2 B(r)} + 
{ (B'(r))^2 \over 4 B^2(r)} + { E^2 \over \hbar^2 B^2(r)} - 
{m_0^2 \over \hbar^2 B(r)} \right] Q(r) = 0
\end{equation}
where $B'$ denotes $dB/dr$ and $B''$ denotes $d^2B/dr^2$.  
Near the horizon, we use the expansion of $B(r)$ given in 
Eqn~(\ref{eqn:horizon}). Neglecting terms of order 
$1/(r-r_0)$ as compared to terms of order $1/(r-r_0)^2$, 
we get, 
\begin{equation}
 -{d^2 Q(r) \over dr^2} - {g \over (r-r_0)^2} Q(r)  = 0  
\qquad {\rm where} \qquad g= \left({1 \over 4} + 
{E^2 \over \hbar^2 R^2}\right)
\label{eqn:sch2}
\end{equation}
Notice that $m_0$ appears nowhere in the above equation.  
Very close to the horizon, the term containing the mass 
does not contribute significantly.  
Eqn~(\ref{eqn:sch2}) is therefore applicable to both massless 
and massive scalar particles.   Making the simple 
transformation $x = (r-r_0)$, we finally obtain 
the effective Schrodinger equation for the system 
with $\hbar =2m=1$ with a potential $(-g/x^2)$. 
\begin{equation}
 -{d^2 Q(x) \over dx^2} - { g  \over x^2} Q(x)  = 0   
\label{eqn:sch3}
\end{equation} 
This potential is symmetric about the origin $x=0$.  
We now include an ``energy'' $\widetilde{E}$ (not to be 
confused with the energy $E$ of the field in the original 
relativistic system) to obtain,   
\begin{equation}
-{d^2 Q(x) \over dx^2} - { g  \over x^2} Q(x)  = 
\widetilde{E}Q(x). 
\label{eqn:sch4}
\end{equation} 
The energy $\widetilde{E}$, which can be positive or 
negative, has been included in order to consider fully 
the properties of the potential $-g/x^2$. 
We will finally be interested in the case $\widetilde{E} =0$ 
which is the case of interest. 
(The energy spectrum is continuous for all values of 
$\widetilde{E}$ which, for $\widetilde{E} < 0$, is surprising 
since for energies less than the potential energy, 
the spectrum is usually discrete.)  
In the subsections to follow we will rederive the semiclassical 
result given in the previous section and then calculate the reflection 
and transmission coefficients using the results of
section~(\ref{subsubsec:potential2}).  
We then look for a correspondence between the reflection and 
transmission coefficients with the semiclassical results. 

\subsubsection{Semiclassical analysis of the 
effective Schrodinger equation}
The semiclassical analysis follows closely the 
method adopted in section~(\ref{sec-hawking}).  
The action functional ${\cal A}$ for a classical particle 
moving in a potential $-g/x^2$ satisfies the 
Hamilton-Jacobi equation
\begin{equation}
{\partial {\cal A} \over \partial t}  + \left( { \partial 
{\cal A} \over \partial x} \right)^2  - {g \over x^2} = 0 
\label{eqn:hjeqn}
\end{equation}
The solution can be immediately written down as,
\begin{equation}
{\cal A} = -\widetilde{E} t \pm \int^x \, {dx \over x} 
\sqrt{\widetilde{E} x^2 + g} \label{eqn:semi1}
\end{equation}
Eqn~(\ref{eqn:semi1}) has an integral which is divergent 
if the action is computed for points lying on the 
opposite sides of the horizon $x=0$.  
Since this has a similar form to Eqn~(\ref{eqn:ssol1}), 
the prescription used in evaluating $S[{\rm emission}]$ 
and $S[{\rm absorption}]$ can be similarly used to evaluate 
${\cal A}[{\rm emission}]$ and ${\cal A}[{\rm absorption}]$.  
The results are 
\begin{eqnarray}
{\cal A}[{\rm emission}]&=& i\pi \sqrt{g} + 
({\rm real \; part}) \nonumber \\
{\cal A}[{\rm absorption}]&=& -i\pi \sqrt{g} + 
({\rm real \; part}). 
\end{eqnarray}
Constructing the semiclassical propagator as before 
and taking the modulus square to obtain the probabilities 
for outgoing and ingoing particles, we get
\begin{equation}
P[{\rm emission}] = \exp\left[ - 4 \pi \sqrt{ {1 \over 4} 
+ { E^2 \over \hbar^2 R^2}} \right] P[{\rm absorption}] . 
\label{eqn:semisys1}
\end{equation}
Note that the result is independent of $\widetilde{E}$.  
Therefore, the above result holds even in 
the case $\widetilde{E}=0$. 
The formula above does not represent a thermal system 
since the energy $E$ appears in the squareroot.  
For large energies such that $(E^2 /\hbar^2 R^2) \gg (1/4)$, 
the above equation can be written in the form 
that suggests a thermal system
\begin{equation}
P[{\rm outgoing}] = \exp\left[ - {4 \pi   E \over \hbar R} 
\right] P[{\rm ingoing}].
\label{eqn:semisys2}
\end{equation}
The temperature $\beta^{-1}$ for the system is the 
same as that in Eqn~(\ref{eqn:stemp}) and one 
recovers the usual result. 
For the case of the Schwarzchild black hole, the condition 
$(E^2 /\hbar^2 R^2) \gg~(1/4)$ reduces to 
$(E/\hbar)\gg 1/4M$. 
In proper units, it can be written as
\begin{equation}
E \gg { \hbar c^3 \over 4 G M} = M_P c^2 {M_P \over 4M} 
\label{eqn:planckcond} 
\end{equation}
where $M_P=\sqrt{\hbar c/G}$ is the Planck mass.  
Since the energy $E$ must be much smaller than the 
Planck energy $M_Pc^2$ in the semiclassical limit, the 
condition above reduces to $M\gg M_P$ i.e. the mass 
of the black hole must be far greater than the Planck mass 
for the spectrum to be thermal.   
This differs significantly from the result given in 
Eqn~(\ref{eqn:ssys}) which does not put any condition 
on the size of the black hole for the spectrum to be thermal.
To verify that the semi-classical analysis is valid, one 
must compute the correction terms and check that these have 
a singularity of the same order as possessed by ${\cal A}$.  
To do this, consider the effective Schrodinger 
equation~(\ref{eqn:sch4}) with factors of $\hbar$ put in.
\begin{equation}
-\hbar^2{d^2 Q(x) \over dx^2} - { g  \over x^2} Q(x)  
= \widetilde{E}Q(x) \label{eqn:ssch4}
\end{equation}
Putting $Q(x) = \exp(iA(x)/\hbar)$, and substituting 
into Eqn~(\ref{eqn:ssch4}), \begin{equation}
-i\hbar{d^2 A(x) \over dx^2} + \left({d A(x) \over dx}\right)^2 
= \widetilde{E} + {g \over x^2}. 
\label{eqn:ssch5}
\end{equation}
Expanding $A$ in powers of $\hbar/i$, we get, 
\begin{equation}
A = {\cal A} + {\hbar \over i} A_1 + 
\left({\hbar \over i}\right)^2 A_2 + \ldots 
\label{eqn:ssch6}
\end{equation}
Substituting into Eqn~(\ref{eqn:ssch5}) and proceeding as usual, 
we find that ${\cal A}$ is given by Eqn~(\ref{eqn:semi1}).  
The next term $A_1$ is given by 
\begin{equation}
A_1 = g \int {dx \over x} {1 \over \widetilde{E}x^2 + g}. \label{eqn:ssch7}
\end{equation}
The relation for $A_1$ also has a singularity at the origin 
of the same order as ${\cal A}$.  Explicit calculation 
of the subsequent terms in the expansion of $A$ reveals 
that all these terms have a singularity of the same order 
as that of ${\cal A}$ and therefore their net contribution 
to the kernel is either as phase factors or as the exponential 
of finite terms multiplied by powers of $\hbar$. 
Therefore, we conclude as before that the 
semiclassical ansatz is valid.

\subsubsection{Tunneling Interpretation}

Instead of the purely semiclassical derivation given 
above, we attempt to see if the relation between the 
reflection and transmission coefficients for the 
effective potential calculated above could be interpreted 
to give a thermal spectrum.  
Using the results in section~(\ref{subsubsec:potential2}) 
and in particular Eqn.~(\ref{eqn:bhpot5}) after setting 
$2m = \hbar = 1$, we obtain,
\begin{equation}
R = T e^{-2\pi\sqrt{g}} = T \exp\left[ - 
{2 \pi   E \over \hbar R} \right] 
\label{eqn:sscht1}
\end{equation}
for ``large'' energies satisfying the inequality 
Eqn.~(\ref{eqn:planckcond}).  
We may interpret the transmission coefficient as 
the probability of emission of particles while the 
reflection coefficient as that for absorption but 
we obtain a temperature which is twice the temperature 
obtained using the semiclassical analysis.  
It is also clear why this is so.  
The  transmission and absorption of particles to and 
from the horizon is clearly not symmetric.  
In the semiclassical analysis, this asymmetry was put 
in by hand in the form of the prescription used to 
evaluate the semiclassical kernel while in the 
tunneling analysis, the potential near the horizon 
is completely symmetric and there is no apriori 
reason for the transmission coefficient to be different 
on either side of the horizon. 
The tunnelling interpretation, in the case of the 
electric field, works in both gauges because of the 
structural similarity of the effective 
Schrodinger equation in both cases.
Further, in the time dependent gauge,  boundary 
conditions play an important role in 
determining particle creation. 
The boundary condition that, at $t = -\infty$, the 
quantum field is initially in a vacuum state, 
introduces an asymmetry into the problem which gives 
rise to particle production in the infinite 
future as $t\to \infty$. 
But, in the space dependent gauge, such a time 
asymmetry in the boundary conditions cannot be 
implemented since the problem is time independent. 
But, since the structure of the effective schrodinger 
equation to be solved is the same as that in 
the time dependent gauge, suitable ratios of the 
transmission and reflection coefficients, which are 
symmetric in both space directions, can be 
identified with the Bogolibov coefficients with 
the subsequent recovery of particle production.  

In the Schwarzchild spacetime, the analogue of the 
``time dependent gauge'' is provided by the study of a 
collapsing shell of matter~\cite{brout95} which, 
close to the black hole horizon, gives rise to production 
of particles which have a thermal distribution in energy. 
In this case,  the boundary condition considered at the 
collapsing shell causes an exponential redshift 
of incoming modes. 
The Bogolibov coefficients can be calculated in a 
straight forward manner and $|\beta_{\lambda}|^2$ is 
Planckian in form and particle production is established 
in the same manner as was done in the electric field case. 
The analogue of the ``space-dependent gauge'' is provided by 
the system where an eternal black hole is in equilibrium with 
a bath of thermal radiation. 
The transmission and reflection coefficients are easily 
determined in this case. 
But, if we attempt to identify the Bogoliubov coefficients 
using ratios of these coefficients, we obtain the result in 
Eqn.~(\ref{eqn:sscht1}) where the tunnelling temperature 
is twice Hawking's result. 
On the other hand, if we use the correspondence 
given in section~(\ref{sec:electric}) for the 
identification we obtain results similar to those obtained 
for the electric field and this does not yield a thermal 
spectrum for the emitted particles.  
A possible qualitative explanation for the factor $2$ 
in the tunnelling temperature is that, since there are 
two disjoint Schwarzchild sectors in the full Kruskal 
manifold, the energy $E$ in  Eqn.~(\ref{eqn:sscht1}) 
should be multiplied by a factor $2$ resulting in the 
recovery of Hawking's result. 
Therefore, the radiation energy seen at infinity is the 
result of tunnelling in both sectors.  
Such an explanation holds for all the spacetimes considered 
in this paper with a horizon, namely, the Rindler and the 
de-Sitter where the full manifold consists of two mutually 
disjoint but identical spacetimes. 
But a more quantitative and satisfactory tunnelling 
interpretation in black hole like spacetimes is needed.

\section{Complex paths interpretation of particle 
production in Electric field} 
\label{sec:cpaths}
In section~(\ref{sec:electric}), we noted how particle 
production can be calculated using 
the tunnelling interpretation.  
This interpretation gives the same result in both the 
space as well as the time dependent gauges. 
The spectrum of particles produced by the electric field 
is not thermal in contrast to the spectrum 
seen by a Rindler observer. 
We use the formal tunnelling method to 
show, in a heuristic manner, how this particle 
production can be obtained by tunnelling between the 
two sectors of the Rindler spacetime.  

The imaginary part of the effective lagrangian 
${\rm Im L}_{\rm eff}$, which is related to the probability 
of the system to remain in the vacuum 
state for all time is given by \cite{paddy91}
\begin{equation}
{\rm Im L}_{\rm eff} = \sum_{n=1}^{\infty} 
\frac{1}{2}\frac{(qE_0)^2}{(2\pi)^3}\frac{(-1)^{n+1}}{n^2}
\exp\left(-\frac{\pi m^2}{qE_0} n\right) 
\label{eqn:efflag}
\end{equation} 
where $m$ is the mass, $q$ is the charge and $E_0$ 
is the magnitude of the electric field. 
We will derive the above expression for 
${\rm Im L}_{\rm eff}$ using the general arguments 
given in section~(\ref{subsec:semiQM}).   

Consider the Hamilton-Jacobi equation for the motion 
of a  particle in an electromagnetic field in $(1+1)$ 
dimensions with a proper time co-ordinate $s$.
\begin{equation} 
\frac{1}{2}\left({\partial F \over \partial t} + qA^t 
\right)^2 - \frac{1}{2}\left({\partial F \over \partial x} 
- qA^x \right)^2 - \frac{1}{2} m^2 + 
\frac{\partial F}{\partial s} = 0 
\label{eqn:cpaths1}
\end{equation}
where $F$ is the action and $A^\mu = (A^t, A^x,0,0)$ 
is the four vector potential. 
We have neglected the dependence of $F$ on 
the $y$ and $z$ co-ordinates.  
This will be justified later on.  
In a time-dependent gauge, given by Eqn.~(\ref{eqn:gauge1}), 
the action $F$ can be easily solved for by using 
the ansatz $F = -Es + p_xx + f(t)$ to give,
\begin{equation}
F (t_1, x_1; t_0,x_0; s) = -Es + p_x(x_1 - x_0) 
\pm \int_{t_0}^{t_1} \! dt\sqrt{(p_x + qE_0t)^2 +
 (m^2 + 2E)} 
\label{eqn:cpaths2t}
\end{equation}
where $p_x$ is the momentum of the particle in the 
$x$ direction and $E$ is the energy of the particle 
corresponding to the proper time co-ordinate $s$.
The trajectory of the particle in the $(t,x)$ plane 
is the usual hyperbolic trajectory given by,
\begin{equation}
(t-t_i)^2 - (x-x_i)^2 = -k_0^2 
\label{eqn:cpaths3}
\end{equation}
where $t_i$, $x_i$ and $k_0$ are suitable constants. 
For any fixed position $x$, there are thus two 
disjoint trajectories corresponding to motion 
in the two Rindler wedges. 

Let us consider the tunnelling of a particle from 
one Rindler trajectory to the other {\it and back} 
in the imaginary time co-ordinate $t$. 
This means that the particle comes back to the same 
spacetime point as it started from. 
Since the process is a tunnelling process, 
the proper time taken is zero. 
Therefore, choosing the positive sign in  
Eqn.~(\ref{eqn:cpaths2t}) (this choice give a 
tunnelling probability that is exponentially damped), 
we have,
\begin{eqnarray}
F(t_0, x_0; t_0,x_0; 0) &=& \oint \! dt
\sqrt{(p_x + qE_0t)^2 + (m^2 + 2E)} = 
{m^2 +2 E\over qE_0} \oint du \sqrt{1 + u^2} 
\nonumber \\
&=& i{m^2 + 2E\over qE_0}\oint d\tau \sqrt{1-\tau^2} = 
i{m^2 + 2E\over qE_0}\int_{0}^{2\pi} d\theta 
\cos^2(\theta) \nonumber \\
&=& {i\pi (m^2 + 2E) \over qE_0} 
\label{eqn:cpaths4}
\end{eqnarray}
where we have made the following changes of variable 
$(p_x + qE_0 t) = u = i\tau$ and $\tau = \sin(\theta)$. 
Taking the limit $E\to 0$, the expression for $\exp(iF)$ 
from the above equation is seen to be exactly the same as 
the  exponential term in Eqn.~(\ref{eqn:efflag}) for $n=1$.
The same argument can be repeated for the particle tunnelling 
$n$ times to and fro to give 
\begin{equation}
F_n(t_0, x_0; t_0,x_0; 0) = i{m^2 + 2E\over qE_0}
\int_{0}^{2n\pi} d\theta \cos^2(\theta) = 
{i\pi (m^2 + 2E) \over qE_0}n . 
\label{eqn:cpaths4a}
\end{equation}
Again, in the limit $E\to 0$, $\exp(iF_n)$ is seen to 
match with the exponential part of the $n$th 
term in  Eqn.~(\ref{eqn:efflag}). 
Therefore, the imaginary part of the total effective 
lagrangian can be written down immediately as
\begin{equation}
{\rm Im L}_{\rm eff} = \sum_{n = 1}^{\infty} 
({\rm prefactor})\, \exp\left(-{\pi m^2 \over qE_0}n\right) 
\label{eqn:cpaths5}
\end{equation}
where the prefactor can only be calculated 
using the exact kernel. 
The dependence of the prefactor on $n$ and the phase 
factor $(-1)^{n+1}$ in Eqn.~(\ref{eqn:efflag}) can 
be deduced using the following arguments. 

The formal expression of the path integral kernel 
for the above electric field problem in the time 
dependent gauge is given by~\cite{paddy91} 
\begin{equation}
K(a,b;s) = \langle a\left| e^{ish} \right|b \rangle 
\label{eqn:cpaths6}
\end{equation}
where $K(a,b;s)$ is the kernel for the particle to 
propagate between the spacetime points $a=(x^0, 
{\bf x})$ and $b=(y^0, {\bf y})$ in a proper time $s$ 
and $h$ is the hamiltonian given by
\begin{equation}
h = \frac{1}{2} (i\partial_i - qA_i)(i\partial^i - qA^i) - 
\frac{1}{2}m^2 
\label{eqn:cpaths7}
\end{equation}
where $A^i$ is the four vector potential given 
in Eqn.~(\ref{eqn:gauge1}) and $q$ and $m$ are the 
charge and mass of the particle respectively. 
Going over to momentum co-ordinates and considering 
the coincidence limit ${\bf x} = {\bf y}$, the kernel 
can be written in the form,
\begin{equation}
K(x^0,y^0; {\bf x},{\bf x}; s) = -\frac{i}
{2 (2\pi)^2} \int_{-\infty}^{\infty}
 \! \frac{dp_x}{s} {\cal G}(x^0, y^0; s) 
\label{eqn:cpaths8}
\end{equation}
where ${\cal G}(x^0, y^0; s)$ is given by
\begin{equation}
{\cal G}(x^0, y^0; s) = \langle x^0 \left| e^{isH} 
\right| y^0\rangle 
\label{eqn:cpaths9}
\end{equation}
and $H$ is the hamiltonian 
\begin{equation}
H = -\frac{1}{2}\left( \frac{\partial^2}{\partial t^2} + 
(p_x + qE_0 t)^2 + m^2 \right) = -\frac{1}{2}
\left( \frac{\partial^2}{\partial \rho^2} + 
q^2E_0^2\rho^2 + m^2 \right) 
\label{eqn:cpaths10}
\end{equation}
with $\rho = (t + p_x/qE_0)$. In the expression for 
the kernel in momentum co-ordinates, we have integrated 
over the transverse momentum variables $p_y$ and $p_z$. 
The above hamiltonian is that of an inverted harmonic oscillator.  
Since all reference to $p_y$ and $p_z$ have disappeared 
in $H$, the dependence of $F$ on the $y$ and 
$z$ co-ordinates was neglected when writing down 
the expression for the Hamilton-Jacobi 
equation in Eqn.~(\ref{eqn:cpaths1}).  
The expression for the effective lagrangian is then 
given by
\begin{equation}
L_{\rm eff} = -i\int_0^{\infty}\! \frac{ds}{s} 
K(x^0,x^0; {\bf x},{\bf x}; s) =  -\frac{1}{2 (2\pi)^2}
\int_0^{\infty}\! \frac{ds}{s^2} 
\int_{-\infty}^{\infty} \! dp_x {\cal G}(x^0, x^0; s) 
\label{eqn:cpaths11}
\end{equation}
We would like to evaluate the propagator 
${\cal G}(x^0, x^0; s)$ for a tunnelling situation 
where the particle tunnels from the 
point $x^0$ and back in loops.  
Since the path integral is not well defined for 
closed loops, it will have to be evaluated in some 
approximate limiting procedure which is outlined below. 

Since the tunnelling potential is that of an inverted 
oscillator, we can use all the results 
of section~(\ref{subsubsec:potential1}). 
The semi-classical wave functions are given in Eqns.~(\ref{eqn:electpot2a},
\ref{eqn:electpot2b}). 
We would like to first account for the 
factor $(-1)^n$ that arises when a particle tunnels 
from one side of the barrier to the other and back. 
 Consider an incident wave to the right of the barrier 
and impinging on it. 
Using the method of complex paths, we rotate this wave 
in the lower complex plane (this is the only route 
possible for the same reason as that given when rotating 
a right moving travelling wave in the upper complex plane) 
to obtain a wave again incident on the barrier with a 
energy independent phase factor $\exp(i\pi/2)$ being 
picked up (other factors dependent on the energy are 
also picked up but these are not important here). 
Since this wave is moving in the wrong direction, 
we assume that the particle that has tunnelled through 
has the same amplitude as the rotated wave but is moving 
away from the barrier.  
This just involves changing the sign of the argument of 
the exponential in the expression for the rotated wave. 
Rotating this left moving wave again in the upper complex 
plane now, the final wave obtained is a right moving wave with 
another extra phase factor of $\exp(i\pi/2)$ being picked up. 
The total phase change with respect to the 
incident wave is thus  $\exp(i\pi)$. 
Since this phase factor is independent of the 
energy, the propagator for the tunnelling process too, 
after one such rotation, will pick up 
a phase of $\exp(i\pi)$.  
Similarly, for $n$ rotations, $n$ taking the values 
$1,2,3,\ldots$, the phase acquired will be 
$\exp(in\pi) = (-1)^n$. 

Therefore, the propagator ${\cal G}$ for $n$ loops, 
${\cal G}_n(x^0, x^0; s)$, can thus be written as
\begin{equation}
{\cal G}_n(x^0, x^0; s) = N(p_x, m, E) e^{in\pi} 
e^{iF_n(x^0, x^0; s)} 
\label{eqn:cpaths13}
\end{equation}
where $F_n(x^0, x^0; s)$ is the classical action for 
the tunnelling process and $N$ is the prefactor 
that arises after evaluating the ``sum over paths''. 
This prefactor is not expected to depend on the proper 
time $s$ since the tunnelling process takes place 
instantaneously or on the number of rotations $n$.  
So the only quantities it may depend on are 
$p_x$, $m$ and $E$. 
Now the expression for the classical action for 
$n$ loops is given in Eqn.~(\ref{eqn:cpaths4a}) 
with $E$ set to zero. 
Substituting this into the expression for the propagator 
in Eqn.~(\ref{eqn:cpaths13}) and evaluating $L_{\rm eff}(n)$ 
for $n$ loops from the relation in Eqn.~(\ref{eqn:cpaths11}) 
we obtain infinity as the answer which is meaningless. 
Therefore, we adopt the following limiting procedure 
in order to obtain a finite and meaningful answer. 
Consider the  action for the tunnelling process.
\begin{equation}
F  = -Es + p_xx \pm \int \! dt\sqrt{(p_x + qE_0t)^2 + 
(m^2 + 2E)} 
\label{eqn:cpaths14} 
\end{equation}
Choosing the positive sign and setting $p_x + qE_0t = 
i\sqrt{m^2 + 2E}\sin(\theta)$, we obtain,
\begin{equation}
F  = -Es + p_xx \pm i\frac{m^2 + 2E}{2qE_0}
\int\! d\theta (1 + \cos(2\theta)) 
\label{eqn:cpaths15}
\end{equation}
For {\it closed} paths only, with $\theta$ taking the 
values from $0$ to $2n\pi$, the above action can be written as,
\begin{equation}
F_n  = -Es \pm i\frac{m^2 + 2E}{2qE_0}\theta 
\label{eqn:cpaths16}
\end{equation}
We have thrown away the $p_x x$ term while retaining the 
$-Es$ term since the dependence on the $x$ 
co-ordinate is really irrelevant.  
Defining a new variable $\bar{\theta} = i\theta/2qE_0$ 
and rescaling $s=\alpha s'$, the above action becomes
\begin{equation}
F_n  = -E\alpha s' + (m^2 + 2E)\bar{\theta} .
\label{eqn:cpaths17}
\end{equation}
Choosing $\alpha$ appropriately, the above action can 
be cast into a form that matches the action for a 
fictitious free particle in $(1+1)$ dimensions with 
``energy'' $\alpha E$ and ``momentum'' $(m^2 + 2E)$ 
satisfying the classical energy-momentum relation,
\begin{equation}
\alpha E = \frac{1}{2}(m^2 + 2E)^2 
\label{eqn:cpaths18}
\end{equation}
where the particle's ``mass'' is set to unity for convenience.  
The above equation determines the quantity $\alpha$. 
Therefore, $F_n$ can also be written in the form
\begin{equation}
F_n (\bar{\theta}_2, \bar{\theta}_1;s) = {(\bar{\theta}_2 -  
\bar{\theta}_1)^2 \over 2 s'} = {\alpha (\bar{\theta}_2 -  
\bar{\theta}_1)^2 \over 2 s} 
\label{eqn:cpaths19}
\end{equation}
where $\bar{\theta}_1$ and $\bar{\theta}_2$ are the 
initial and final states of the free particle with 
$s'$ being the proper time taken. (Note that 
$(\bar{\theta}_2 -  \bar{\theta}_1) = 2in\pi/2qE_0$.)  
Substituting this into the expression for 
${\cal G}_n(x^0, x^0; s)$ in Eqn.~(\ref{eqn:cpaths13}) 
and evaluating only the integral over $s$ in the 
expression for the effective lagrangian in 
Eqn.~(\ref{eqn:cpaths11}) {\it without} 
taking the limits, we obtain
\begin{eqnarray}
\int\! \frac{ds}{s^2}{\cal G}_n(x^0, x^0; s) &=& 
N(p_x, m, E) e^{in\pi} \int \! \frac{ds}{s^2} 
\exp\left({i\alpha (\bar{\theta}_2 -  
\bar{\theta}_1)^2 \over 2 s}\right) \nonumber \\
&=& -N(p_x, m, E) e^{in\pi} \frac{2}{i\alpha 
(\bar{\theta}_2 -  \bar{\theta}_1)^2} 
\exp\left({iF_n(\bar{\theta}_2, \bar{\theta}_1;s)}\right) .
\label{eqn:cpaths20}
\end{eqnarray}
Notice that the prefactor to the exponential has no 
dependence on the proper time $s$. 
Now, we use the form for the action given by 
Eqn.~(\ref{eqn:cpaths16}) and taking the limits 
for $s$ from $0$ to $\infty$, the effective 
lagrangian for $n$ loops, $L_{\rm eff} (n)$ 
can be written in the form, 
\begin{equation}
L_{\rm eff} (n) = \frac{i}{2} \frac{(qE_0)^2}{(2\pi)^3} 
\frac{(-1)^{n+1}}{n^2} \exp\left(-\frac{\pi m^2}
{qE_0} n\right) \frac{8}{\alpha} 
\exp\left(-\frac{2\pi E}{qE_0} n\right) 
\int_{-\infty}^{\infty} \! \frac{dp_x}{2\pi} N(p_x, m, E)  
\label{eqn:cpaths21}
\end{equation}
where we have set $(\bar{\theta}_2 -  \bar{\theta}_1) 
= 2in\pi/2qE_0$. 
Taking the limit of $E\to 0$ and using the 
expression for $\alpha$ in Eqn.~(\ref{eqn:cpaths18}), 
we find that $N$ must satisfy the relation
\begin{equation}
\lim_{E\to 0} \frac{16E}{(m^2 + 2E)^2} \,
\int_{-\infty}^{\infty} \! \frac{dp_x}{2\pi} 
N(p_x, m, E)  = 1 
\label{eqn:cpaths22}
\end{equation}
so that the imaginary part of the effective 
lagrangian for $n$ loops, $L_{\rm eff}(n)$, 
matches the $n$th term in Eqn.~(\ref{eqn:efflag}). 
Therefore, in this manner, the contributions to 
the imaginary part of the effective lagrangian 
for the uniform electric field can be thought of 
as arising from the tunnelling of particles 
between the two Rindler sectors.

\section{Conclusions}
In conclusion, we see that the tunnelling picture 
is not a valid picture in general.  
In the case of the electric field, it can successfully 
be applied to both the time dependent and time 
independent gauges but such a picture is valid only 
to a limited extent. 
When applied naively to black hole like spacetimes it 
gives a temperature that is twice the expected value. 
However, by taking into account the presence of 
the two disjoint Schwarzchild sectors in the full 
Kruskal manifold this discrepancy can be corrected. 
The semi-classical prescription given in the paper 
takes into account the asymmetry between the two 
sides of the horizon and gives the correct result 
without requiring the Kruskal extension. 
An interesting aspect in the reduction to the 
effective Schrodinger problem is that the 
semi-classical analysis after the reduction does 
not give a thermal spectrum for all energies.  
Only when the mass of the black hole is much greater 
than the Planck mass does the spectrum reduce 
to a thermal form.    

But the tunnelling picture does seem to give a 
nice interpretation of particle production in an 
electric field as arising due to tunnelling between 
the two disjoint sectors of the Rindler spacetime. 
Though we have only given a heuristic argument in 
this paper, we will explore this issue further 
in a future publication.  

\section*{Acknowledgements}
\noindent
KS is being supported by the Senior Research Fellowships 
of the Council of Scientific and Industrial Research, India.

\section*{\centerline{Appendix:~Generalization to 
(3+1)--dimensions}}
\label{sec-3d}
The generalization to (3+1) dimensions is straightforward. 
In this section, the semi-classical analysis given in
 section~(\ref{sec-hawking}) will be briefly outlined.  
Further, it will be shown that the effective Schrodinger 
equation in (3+1) dimensions is the same 
as Eqn~(\ref{eqn:sch3}).

Consider the metrics (\ref{eqn:metric2}) and (\ref{eqn:metric3}).  
We will work with  (\ref{eqn:metric2}) which is in 
spherical polar co-ordinates. 
The results obtained are extendable to (\ref{eqn:metric3}) 
in a straightforward manner.  

The Klein-Gordon equation, written using the 
metric~(\ref{eqn:metric2}), is 
\begin{eqnarray}
{1 \over B(r)} {\partial^2 \Phi \over \partial t^2} - 
{1 \over r^2}{\partial \over \partial r} 
\left( r^2 B(r) {\partial \Phi \over \partial r}\right) 
- && {1 \over r^2 \sin(\theta)}{\partial \over 
\partial \theta}\left( \sin(\theta) {\partial \Phi 
\over \theta} \right) \nonumber \\
&& - {1 \over r^2 \sin^2(\theta)}{\partial^2 
\Phi \over \partial \phi^2}
  = -{m_0^2 \over \hbar^2} \Phi 
\label{eqn:kgr3}
\end{eqnarray}
Since the problem is a spherically symmetric one, one 
can put $\Phi = \Psi(r,t) Y^m_l(\theta, \phi)$ to obtain,
\begin{equation}
{1 \over B(r)} {\partial^2 \Psi \over \partial t^2}  - 
{1 \over r^2}{\partial \over \partial r} 
\left( r^2 B(r) {\partial \Psi \over \partial r}\right) 
+ \left( {l(l+1) \over r^2} + {m_0^2 \over \hbar^2} 
\right)\Psi = 0
\end{equation}
Making the ansatz $\Psi = \exp((i/\hbar) S(r,t))$ 
and substituting into the above equation, we obtain,
\begin{eqnarray}
& &\left[ {1 \over B(r)}\left({\partial S \over 
\partial t}\right)^2 - B(r)\left({\partial S \over 
\partial r}\right)^2 -m_0^2 - {l(l+1) \hbar^2 
\over r^2} \right] \nonumber \\
&& \quad + {\hbar \over i} \left[ {1 \over B(r)}
{\partial^2 S \over \partial t^2} - B(r) 
{\partial^2 S \over \partial r^2} - {1 \over r^2}
{d (r^2 B) \over dr} {\partial S \over \partial r} 
\right] = 0
\end{eqnarray}
Expanding $S$ in a power series as in 
Eqn~(\ref{eqn:exp}), we obtain, to the 
zeroth order in $\hbar/i$, 
\begin{equation}
S_0 = - Et \pm \int^r {dr \over B(r)} 
\sqrt{E^2 - B(r)( m_0^2 + L^2/r^2)} 
\label{eqn:3ssol1}
\end{equation}
where $L^2 = l(l+1)\hbar^2$ is the angular momentum.  
It is easy to see from the above equation that near 
the horizon, the presence of the $L^2$ term can 
be neglected since it is multiplied by $B(r)$. 
Therefore, the  semiclassical result of 
section~(\ref{sec-hawking}) follows even in the 
case of (3+1)--dimensions. 
The semi-classical ansatz is valid in this case as 
can be seen by calculating explictly the higher 
order terms in the expansion for $S$.  
All these terms have a singularity at the horizon 
of the same order as that of $S_0$ and they contribute 
to the semiclassical propagator either as phase factors 
or as terms multiplied by powers of $\hbar$ 
which are entirely negligible.  
Expanding the Klien-Gordon equation for 
$\Phi$ using the metric~(\ref{eqn:metric3}) 
gives analogous results and will 
not be explictly given.  

Now, consider the reduction of Eqn~(\ref{eqn:kgr3}) 
to an effective Schrodinger equation. 
Setting 
\begin{equation}
\Psi = \exp(-iEt/\hbar) Y^m_l(\theta,\phi) 
\Psi(r) 
\end{equation}
and substituting into Eqn~(\ref{eqn:kgr3}), we obtain,
\begin{equation}
B(r){d^2 \Psi \over dr^2} + {1 \over r^2}{d(r^2B) 
\over dr}{d \Psi \over dr} + \left( {E^2 \over \hbar^2 B(r)} 
- {m^2 \over \hbar^2} - {L^2 \over \hbar^2 r^2} \right) 
\Psi = 0
\end{equation}
Making the substitution
\begin{equation}
\Psi = {1 \over \sqrt{r^2 B(r)}} Q(r)
\end{equation}
we get the result,
\begin{equation}
-{d^2 Q \over dr^2} - \left[ {1 \over B^2}
\left( {(B')^2 \over 4} + {E^2 \over \hbar^2}\right) 
- {1 \over B}\left( {B'' \over 2} +{B'\over r} + 
{m_0^2 \over \hbar^2} + {L^2 \over \hbar^2 r^2}\right) 
\right] Q = 0
\end{equation}
where $B'$ and $B''$ are the first and second 
derivatives of $B(r)$ respectively.  
Near the horizon $r = r_0$, using the expansion 
for $B(r)$ given in Eqn~({\ref{eqn:horizon}), 
it is easy to see that the $1/B^2$ term in the 
above equation dominates over the $1/B$ term which 
can therefore be neglected.  
The resulting Schrodinger equation is the same 
as in Eqn~(\ref{eqn:sch3}).  
Therefore, even in (3+1) dimensions, the 
semiclassical and quantum mechanical 
results of section~(\ref{sec:schrodinger}) 
are the same.  
It can be easily proved that the effective 
Schrodinger equation for the cartesian 
metric~(\ref{eqn:metric3}) gives the same result.

\end{document}